\documentclass[twocolumn]{aastex701}

\usepackage{stackengine}
\usepackage{amsmath}
\usepackage{subfig}
\usepackage[percent]{overpic}
\usepackage[dvipsnames]{xcolor}
\usepackage{soul}
\usepackage{xcolor}

\begin{document}

\title{CHANG-ES \uppercase\expandafter{\romannumeral 40}: Magnetic Field Structures in the Disk and Halo of NGC~891}

\author[orcid=0009-0006-5216-747X]{N.Pourjafari}
\affiliation{Department of Physics and Astronomy, The University of Calgary,
2500 University Drive NW,
Calgary AB, T2N 1N4, Canada} 
\email{niloofar.pourjafari@ucalgary.ca}

\author[orcid=0000-0003-2623-2064]{J. M. Stil} 
\affiliation{Department of Physics and Astronomy, The University of Calgary,
 2500 University Drive NW,
Calgary AB, T2N 1N4, Canada}
\email{jstil@ucalgary.ca}

\author[orcid=0000-0001-8206-5956]{R.-J. Dettmar}
\affiliation{Ruhr University Bochum, Faculty of Physics and Astronomy, Astronomical Institute (AIRUB), 44780 Bochum, Germany}
\email{dettmar@astro.ruhr-uni-bochum.de}

\author[orcid=0000-0002-5425-6074]{P. Kamphuis}
\affiliation{Ruhr University Bochum, Faculty of Physics and Astronomy, Astronomical Institute (AIRUB), 44780 Bochum, Germany}
\affiliation{Currently at Instituto de Astrof\'isica de Andaluc\'ia, Glorieta de Astronom\'ia, Granada, IAA-CSIC, Spain}
\email{peterkamphuisastronomy@gmail.com}

\author[orcid=0009-0001-8154-3562]{R. Beck}
\affiliation{Max Planck Institute for Radio Astronomy, Bonn, Germany}
\email{rbeck@mpifr-bonn.mpg.de}

\author[orcid=0000-0001-5310-1022]{J. English}
\affiliation{University of Manitoba, Department of Physics and Astronomy, Winnipeg, Manitoba, R3T 2N2, Canada}
\email{jayanne.english@umanitoba.ca}

\author[orcid=0000-0002-2082-407X]{V. Heesen}
\affiliation{Hamburger Sternwarte, University of Hamburg, Gojenbergsweg 112, 21029, Hamburg, Germany}
\email{volker.heesen@uni-hamburg.de}

\author[orcid=0000-0003-0073-0903]{J. Irwin}
\affiliation{Department of Physics, Engineering Physics and Astronomy, Queen's University, Kingston, ON, K7L 3N6, Canada}
\email{irwinja@queensu.ca}

\author[orcid=0000-0001-6239-3821]{J.-T. Li}
\affiliation{Purple Mountain Observatory, Chinese Academy of Sciences, 10 Yuanhua Road, Nanjing, 210023, China}
\email{pandataotao@gmail.com}

\author[orcid=0000-0002-3286-5346]{L.-Y. Lu}
\affiliation{Department of Physics and Astronomy and Research Center of Astronomy, Qinghai University, 251 Ningda Road, Xining, 810016,
China}
\email{luliyuan@stu.xmu.edu.cn}

\author[orcid=0000-0002-9559-3827]{S. Ranasinghe}
\affiliation{Department of Physics and Astronomy, The University of Calgary,
2500 University Drive NW, Calgary AB, T2N 1N4, Canada}
\email{syran3@hotmail.com}

\author[orcid=0000-0001-8428-7085]{M. Stein}
\affiliation{Ruhr University Bochum, Faculty of Physics and Astronomy, Astronomical Institute (AIRUB), 44780 Bochum, Germany}
\email{mstein@astro.ruhr-uni-bochum.de}

\author[orcid=0000-0002-9279-4041]{Q. D. Wang}
\affiliation{Astronomy Department, University of Massachusetts, Amherst, MA 01003, USA}
\email{wqd@umass.edu}

\author[orcid=0000-0002-3502-4833]{T. Wiegert}
\affiliation{Instituto de Astrofísica de Andalucía (IAA-CSIC), Glorieta de la Astronomía, 18008, Granada, Spain}
\email{theresa.wiegert@gmail.com}

\begin{abstract}

We present new Karl G. Jansky Very Large Array S-band (2-4 GHz) observations of the edge-on spiral galaxy NGC~891, complemented by C-band data, to investigate the structure of its radio continuum halo. Using rotation measure synthesis we detected an extended polarized halo, with most spatially extended polarized emission confined to Faraday depths within $\pm 150\ \rm rad\ m^{-2}$. We identified a localized region in the north-east side of the galaxy that shows an enhancement in polarized intensity (not in percentage polarization). By combining the radio data with $\rm H\alpha$ and diffuse X-ray maps, we discuss a possible origin for this structure: a superbubble powered by clustered supernovae. Across the disk and halo, the percentage polarization decreases toward the midplane but shows a mild wavelength dependence, despite the edge-on orientation of NGC~891. This behavior implies that the depolarization cannot be dominated by small-scale Faraday rotation within the disk. Instead, it is possible that most of the observed polarized emission arises on the Earth-facing side of the galaxy. Our peak rotation measure (RM) map shows a smooth transition along the major axis, consistent with a large scale axisymmetric magnetic field.
Using $\rm H\alpha$ and UV data, we analyzed the distribution of H\,{\footnotesize II} regions and found that they are parts of different spiral arms.
We also identified a faint, isolated H\,{\footnotesize II} region at a galactocentric radius of $16.9\ \rm kpc$, with both $\rm H\alpha$ and far-UV counterparts, indicating star formation outside the thin disk.

\end{abstract}


\section{Introduction}

NGC~891 is a nearby edge-on spiral galaxy located at $RA\ (J2000) = 02^{\rm h}22^{\rm m}32\fs91$ , $DEC\ (J2000) = +42\arcdeg20\arcmin 53\farcs95$ \citep{van2012radio}, and its distance is $9.1\ \rm Mpc$ \citep{wiegert2015chang, radburn2011ghosts} with an inclination of $89\fdg8$ \citep{kregel2005structure} and a Hubble type of $\rm Sb$ 
\citep[From HYPERLEDA\footnote{\url{http://leda.univ-lyon1.fr/}},][]{paturel2003hyperleda,makarov2014}. 
It has long been considered a close analog, or “twin” of the Milky Way owing to its similar optical luminosity \citep{de1991book}, morphology \citep{van1981surface}, and rotation velocity \citep{rupen1991neutral, oosterloo2007cold, fraternali2011tale}. However, compared to the Milky Way, NGC~891 exhibits a somewhat higher star formation rate \citep{arshakian2011modeling}, consistent with its larger reservoir of molecular gas while maintaining a radial CO distribution  similar to that of the Milky Way \citep{scoville1993scale}. Its proximity and almost perfectly edge-on orientation make NGC~891 one of the best-studied spiral galaxies across the electromagnetic spectrum, from radio \citep[e.g.][]{dahlem1994spatially, irwin2012continuum, wiegert2015chang, mulcahy2018investigation, schmidt2019chang, krause2020chang} to X-rays \citep[e.g.][]{bregman1994x, hodges2012deep}.

The density of Warm Ionized Medium (WIM) in NGC~891 is approximately twice the density of the WIM in the Milky Way at the same galactocentric radius \citep{rand1990}. Deep H$\alpha$ images show shells and vertical filaments with [S\,{\footnotesize II}]/H$\alpha$ ratio consistent with photoionization \citep{rand1990}. 
High-resolution imaging with the Hubble Space Telescope by \citet{rossa2004hubble} shows a bright and vertically extended diffuse ionized gas (DIG) layer reaching well into the halo with vertical filaments and supershells. 

Therefore, NGC~891 is regarded as a benchmark system for investigating disk–halo interactions and the structure of galactic halos. While our own Galaxy can be studied in more detail, its global properties are difficult to disentangle from our embedded vantage point. NGC~891, as a nearby, nearly edge-on, Milky Way analog, provides the ideal opportunity for investigating the structure of a galactic halo, and the disk-halo interactions.

Radio continuum emission, particularly its polarized component, provides one of the most powerful probes of magnetic fields in spiral galaxies. Radio continuum studies of NGC~891 span several decades and have revealed the presence of both disk and halo components in its emission \citep{allen1978radio}. 
Early polarization studies at centimeter wavelengths showed large-scale ordered magnetic field patterns in the disk and halo \citep[e.g.][]{sukumar1991}, including halo emission at lower frequencies \citep{hummel1991}, and presented models of depolarization by internal Faraday dispersion in the halo \citep{hummel1991b}. Low-frequency observations with LOFAR have provided the high-sensitivity images of NGC 891 at meter wavelengths and constrained the vertical scale heights of its radio halo \citep{mulcahy2018investigation}.

Observations of edge-on spirals, show plane-parallel magnetic fields within the disk \citep[e.g.][]{dumke1995polarized,krause2020chang}, but reveal an “X-shaped” geometry in the halo \citep[e.g.][]{golla1994intrinsic, tullmann2000thermal, krause2006large,krause2008magnetic, stein2025chang}. The magnetic field strength in these halos is often comparable to that in the underlying disk \citep{krause2019magnetic}.

CHANG-ES (Continuum Halos in Nearby Galaxies - an EVLA Survey) is a survey of radio continuum and polarization of 35 nearby nearly edge-on spiral galaxies with the Karl G. Jansky Very Large Array (VLA).

In this paper we present and analyze the first radio observations of NGC~891 in S-band (2-4 GHz), observed as part of the CHANG-ES survey with the VLA array in C configuration. S-band represents an important wavelength range in which most of the transition from moderate to strong depolarization occurs in star forming galaxies. Previously, CHANG-ES observed NGC~891 in L-band \citep[][Walterbos et al. in prep.]{schmidt2019chang, stein2023} and in C-band \citep{irwin2012continuum, wiegert2015chang}. In L-band, linear polarization was detected from a few regions in the northeast \citep{schmidt2019chang}, while in C-band widespread diffuse polarized emission is detected \citep{krause2020chang}. In this paper we present new S-band observations, and analyze the linear polarization jointly in S-band and C-band for maximum sensitivity to Faraday thick structures. Including L-band is deferred to future work because of the bright radio source 3C66 near the edge of the primary beam \citep{mulcahy2018investigation, schmidt2019chang}, and because the lack of strong polarized emission in L-band from most parts of the galaxy.    

The remainder of this paper is organized as follows: In Section \ref{sec:Data_and_methods} we describe the observational setup and data processing pipeline, with subsections covering the specifics of our observations and data processing, polarization calibration, imaging techniques, and the RM synthesis methodology used to analyze the radio data, as well as ancillary data from other sources. Section \ref{sec:Results} presents our findings, including total intensity images and RM synthesis and other wavelengths results. In Section \ref{sec:Discussion} we delve deeper into the interpretation of our data. Finally, Section \ref{sec:Conclusion} summarizes our key findings. Throughout this paper, all coordinates are given in the J2000 reference frame.

\section{Data and methods} \label{sec:Data_and_methods}
\subsection{Observations and data processing}
Two observations were made for the S-band data: June 29, 2021 and August 13, 2021. In each case, 3.13 hours on source were obtained. The observing information is summarized in Table \ref{tab:deluxesplit}. The observations of NGC~891 were carried out with two pointings, centered at $RA = 2^{\rm h}22^{\rm m}37\fs21$, $DEC = +42\arcdeg 22 \arcmin 41\farcs2$ and $RA = 2^{\rm h}22^{\rm m}29\fs61$, $DEC = +42\arcdeg 19 \arcmin 12\farcs6$, in interleaved mode.
3C48 was used as the primary flux calibrator with a total flux density at $3\ \rm GHz$ of $8.5\ \rm Jy$ \citep{perley2017accurate} and as polarization angle calibrator along with 3C84 (Perseus A) as unpolarized calibrator. The source J0251+4315 was used as the phase calibrator. Polarization calibration and imaging was done in CASA (Common Astronomy Software Applications) environment 6.5.4 \citep{mcmullin2007casa}. A single phase self-calibration loop was applied on the first observation, followed by two amplitude-phase self-calibration loops. There was little improvement after the initial phase only self-calibration. Symmetric clean residuals in the second observation prompted us to start with amplitude and phase calibration. The data were manually flagged to remove any radio frequency interference (RFI). 

\begin{figure}
    \centering

    \begin{overpic}[trim=20 20 20 10, clip, width=0.95\columnwidth]{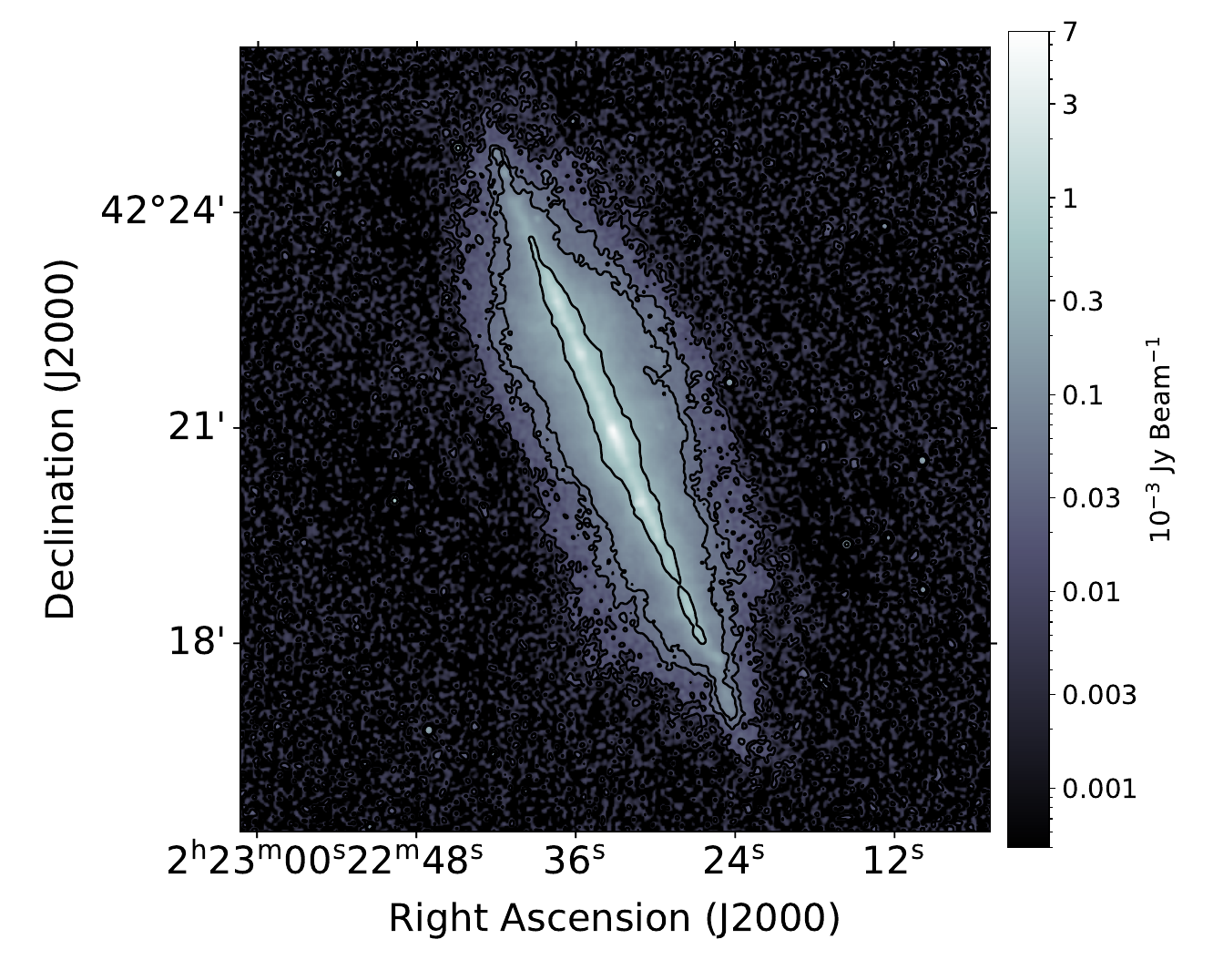}
        \put(75,73){\color{black}\setlength\fboxsep{1pt}\fcolorbox{black}{white}{\textbf{(a)}}}
    \end{overpic}
    \vspace{2mm}

    \begin{overpic}[trim=20 20 15 10, clip, width=0.95\columnwidth]{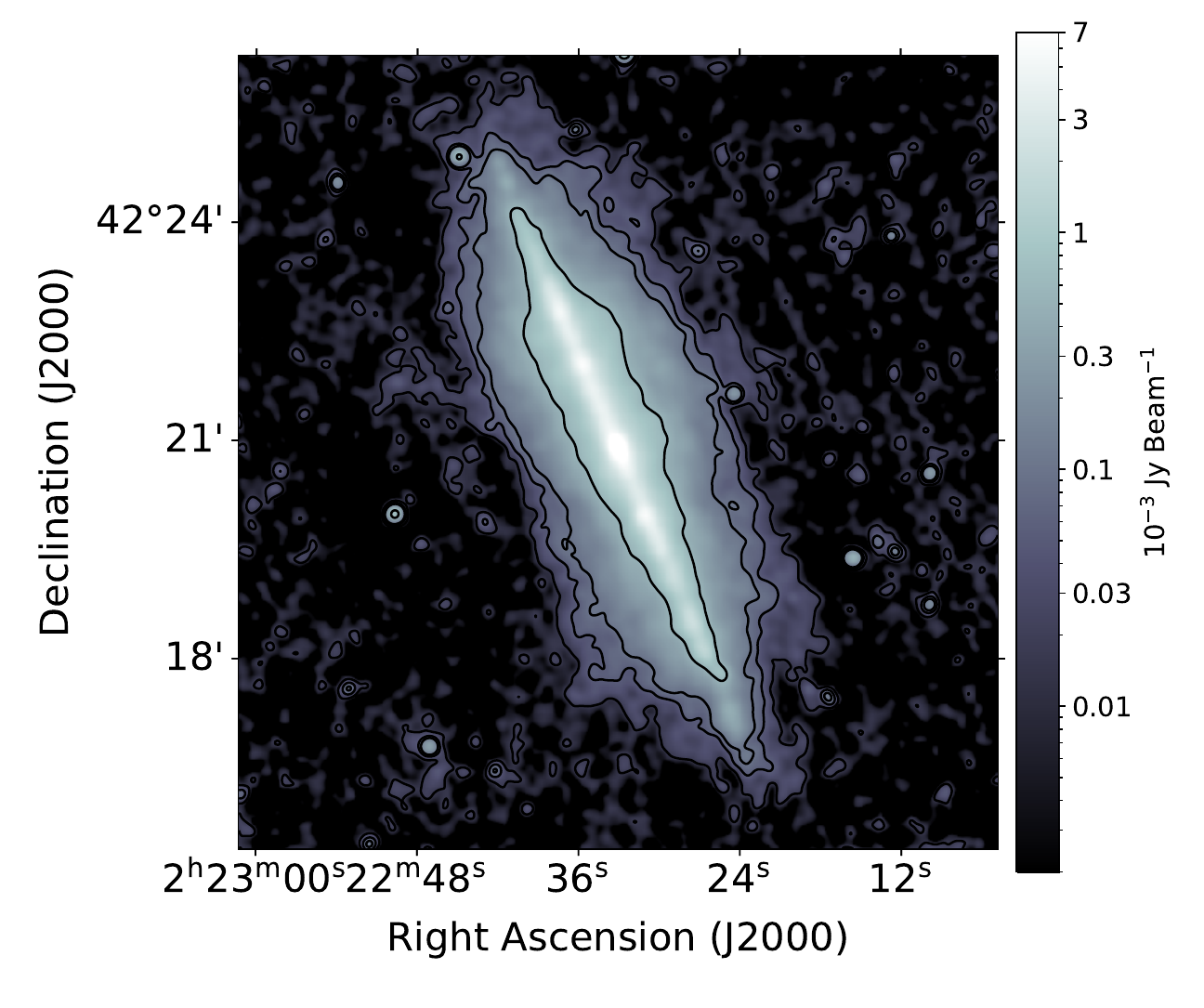}
        \put(75,75){\color{black}\setlength\fboxsep{1pt}\fcolorbox{black}{white}{\textbf{(b)}}}
    \end{overpic}
    \vspace{2mm}

    \begin{overpic}[trim=20 20 20 10, clip, width=0.95\columnwidth]{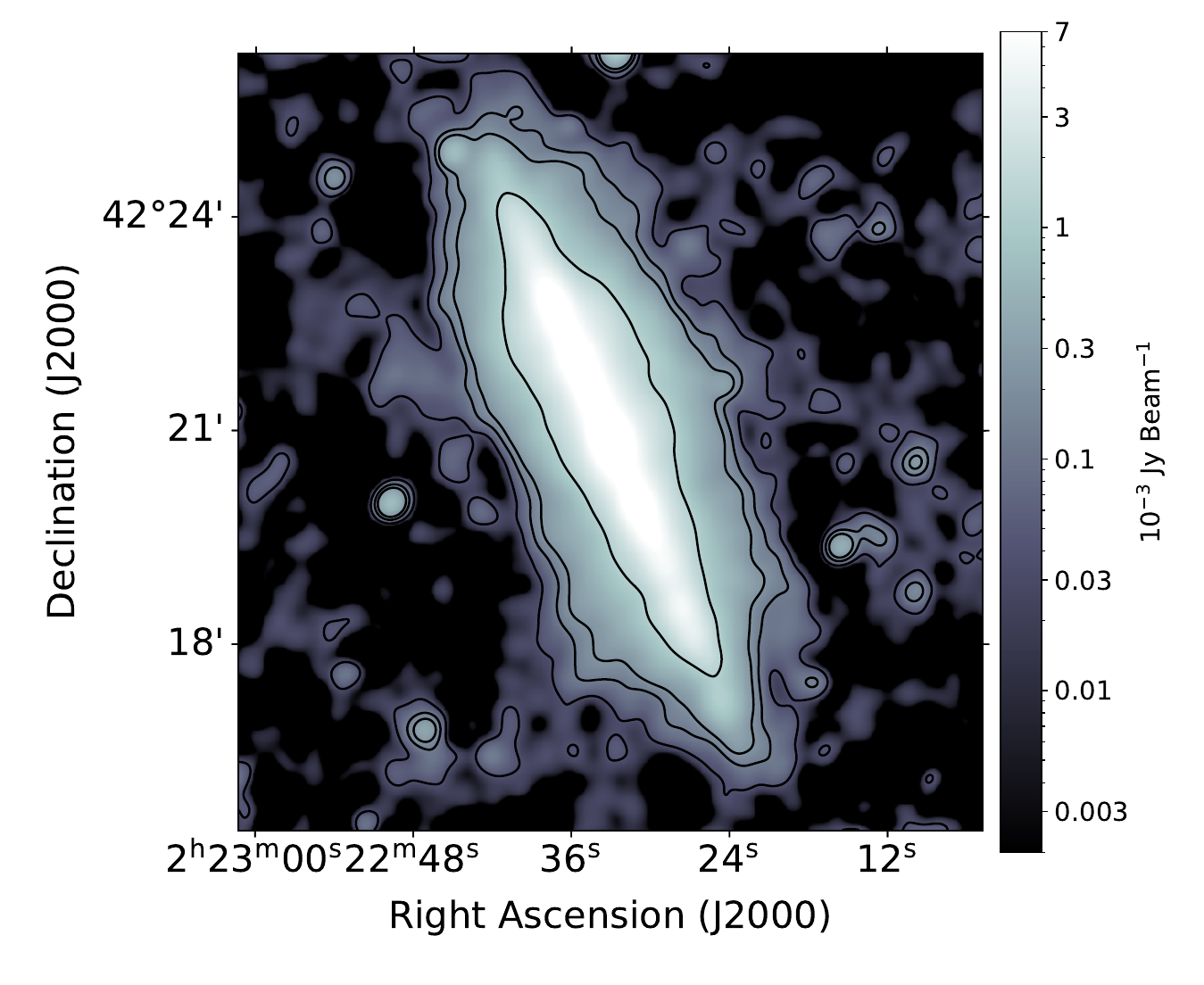}
        \put(75,73){\color{black}\setlength\fboxsep{1pt}\fcolorbox{black}{white}{\textbf{(c)}}}
    \end{overpic}

    \caption{Primary beam corrected S-band total intensity images of NGC~891 at a resolution of a: $4\farcs92 \times 4\farcs65$ FWHM (rms noise: $\sigma_I = 3.4\ \rm \mu Jy\ Beam^{-1}$), b: $10\farcs10 \times 9\farcs83$ FWHM (rms noise: $\sigma_I = 5.4\ \rm \mu Jy\ Beam^{-1}$), c: $20\farcs65 \times 18\farcs83$ FWHM (rms noise: $\sigma_I = 12.5\ \rm \mu Jy\ Beam^{-1}$). The Stokes $I$ contour levels are: $3 \sigma_I, 10 \sigma_I, 18 \sigma_I, 100 \sigma_I$. The images are displayed using a logarithmic intensity scale. The different angular resolutions were obtained using uv-tapering (see Table~\ref{tab:3maps} for details).}
    \label{fig:Stokes_I_images}
\end{figure}

\begin{deluxetable}{lc}
\tabletypesize{\scriptsize}
\tablewidth{0pt}
\tablecaption{Parameters of observations of NGC 891 in S band
\label{tab:deluxesplit}}
\startdata
\\[0.5mm]
Frequency band (GHz) & 2-4 \\ Configuration & C \\ No. of antennas
& 26 \\ No. of spws$^a$ & 16 \\ No. of channels/spw & 64 \\ Flux and bandpass calibrator & 3C48 \\ Phase
calibrator & J0251+4315 \\ Zero polarization calibrator & 3C84 \\
\enddata
\tablecomments{$^a$ spw refers to 'spectral window', such that 16 spws across 2 GHz of total bandwidth corresponds to 125 MHz per spw.}
\end{deluxetable}

\begin{figure*}
    \centering
    \includegraphics[trim=370 30 320 80, clip, width=\textwidth]{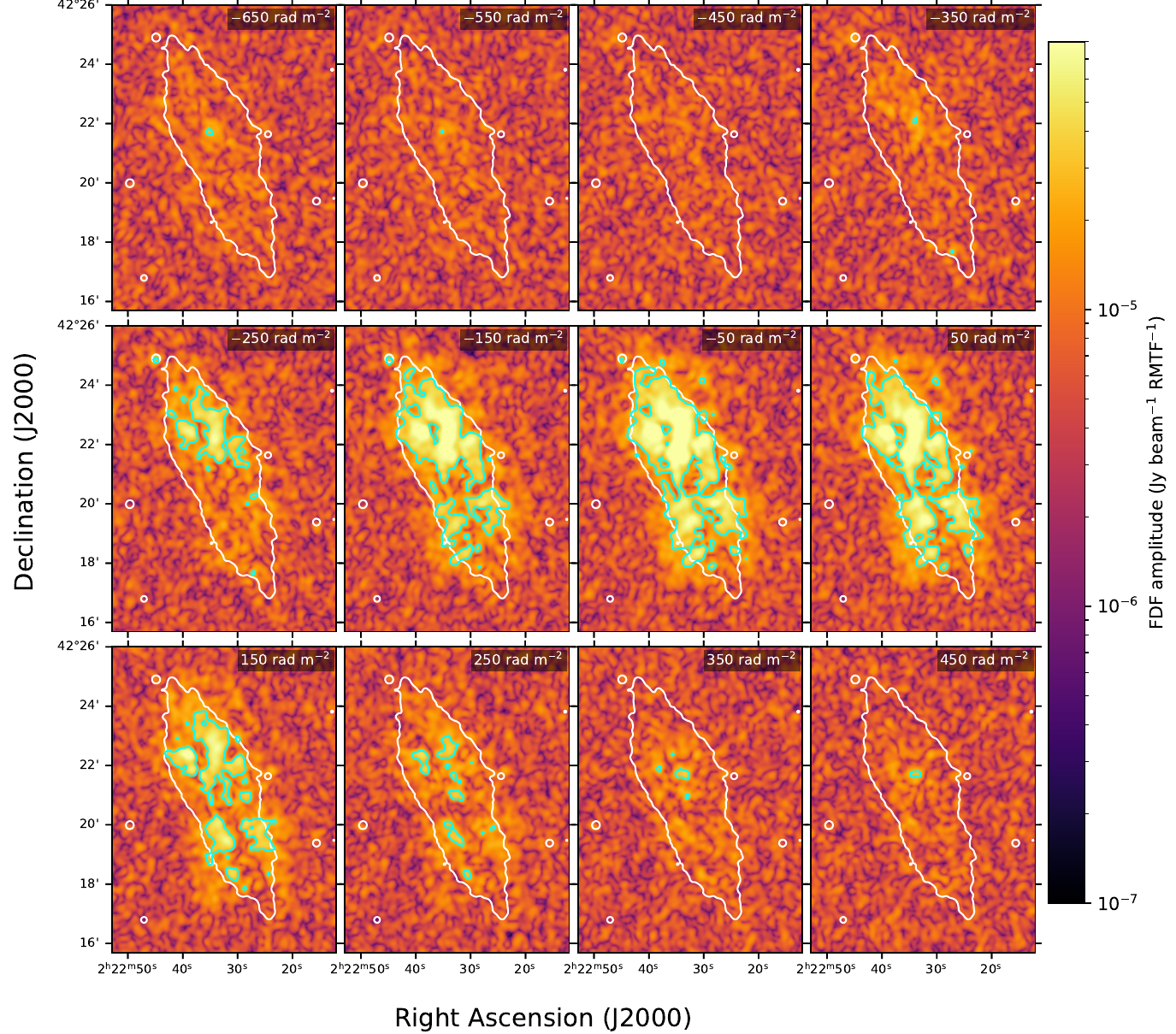}
    \caption{Results of the three-dimensional RM synthesis and RM clean procedures for NGC 891. White contours indicate Stokes $I$ emission at the $18 \sigma_I$ level ( $\sigma_I = 5.4\ \rm \mu Jy\ Beam^{-1}$). Cyan contours mark regions of polarized intensity above the $8 \rm \sigma$ detection threshold, where $\rm \sigma = 3.3\times10^{-6}\ \rm Jy\ Beam^{-1}\ RMTF^{-1}$.}
    \label{fig:ngc891_3d_rm_synthesis_grid}
\end{figure*}

\subsection{Polarization Calibration}
Polarization calibration was performed in three main steps: determining the instrumental delay between the two polarization outputs, solving for the instrumental polarization, and solving for the polarization position angle using 3C48. The overall calibration procedure followed the approach described by \citet{irwin2012continuum} with modifications to account for the frequency-dependent fitting of the S-band polarization model. We fitted these curves to the S-band data for the percentage polarization ($p_\nu$) and the polarization angle ($\chi_\nu$) of 3C48:

\begin{equation}
\begin{split}
    p_\nu &= 0.003291(\frac{\nu -3 }{3})^3 - 0.02093(\frac{\nu - 3}{3})^2 \\
    & + 0.04539(\frac{\nu - 3}{3}) + 0.02204
\end{split}
\end{equation}

\begin{equation}
\begin{split}
    \chi_\nu &= -0.001384(\frac{\nu - 3}{3})^4 +0.02117(\frac{\nu-3}{3})^3 \\
    & -0.1290(\frac{\nu-3}{3})^2 + 0.3451(\frac{\nu-3}{3}) \\
    & -1.420
\end{split}
\end{equation}

where $\nu$ is in GHz.

\begin{figure}
    \centering

    \begin{overpic}[trim=70 0 100 70, clip, width=0.95\columnwidth]{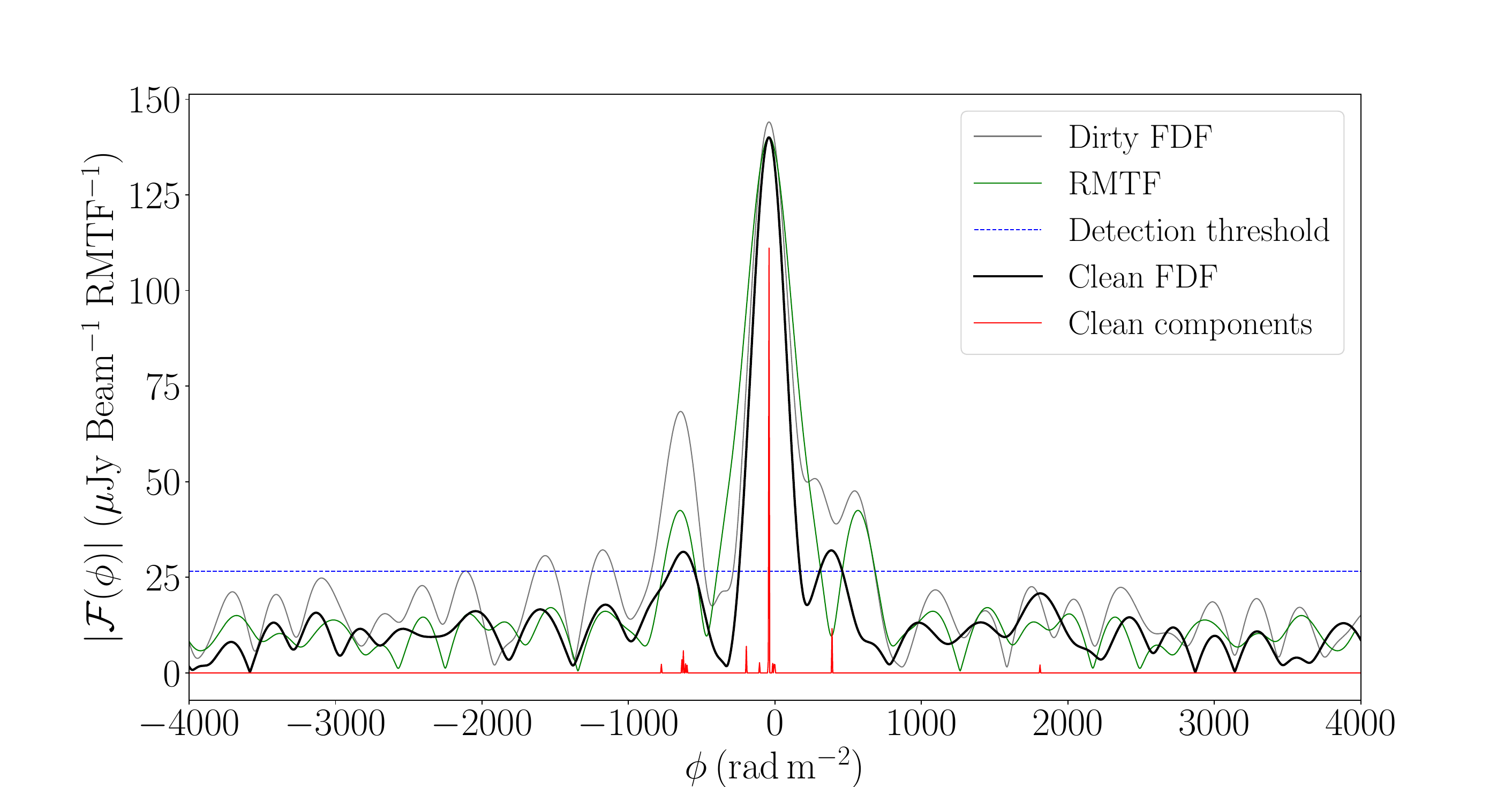}
        \put(10,46){\color{black}\setlength\fboxsep{1pt}\fcolorbox{black}{white}{\textbf{(a)}}}
    \end{overpic}

    \begin{overpic}[trim=10 30 5 20, clip, width=0.95\columnwidth]{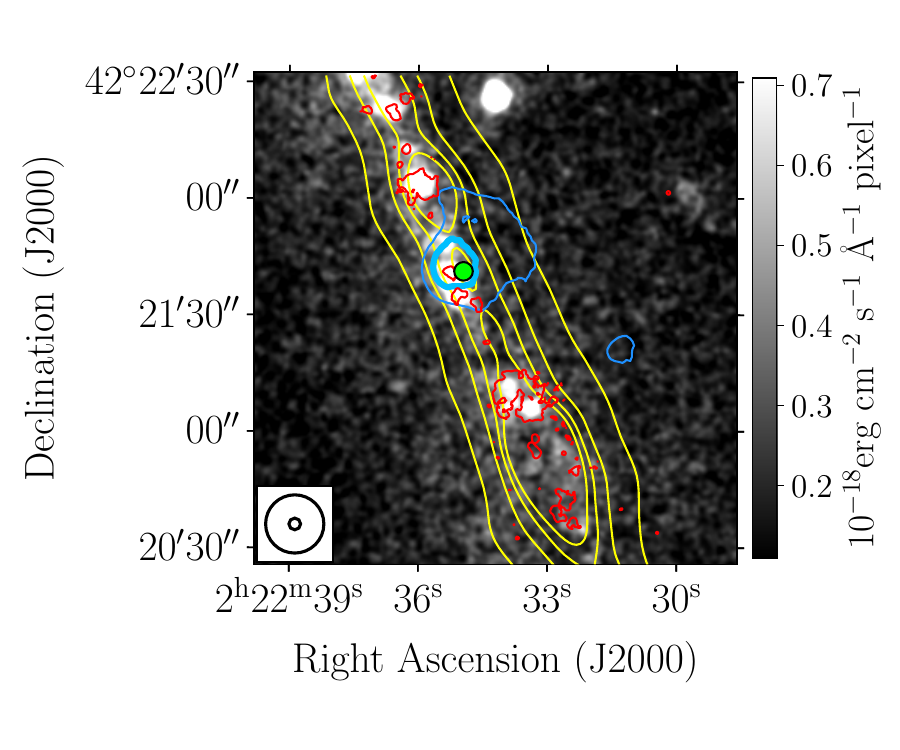}
        \put(28,67){\color{black}\setlength\fboxsep{1pt}\fcolorbox{black}{white}{\textbf{(b)}}}
    \end{overpic}

    \begin{overpic}[trim=10 30 5 20, clip, width=0.95\columnwidth]{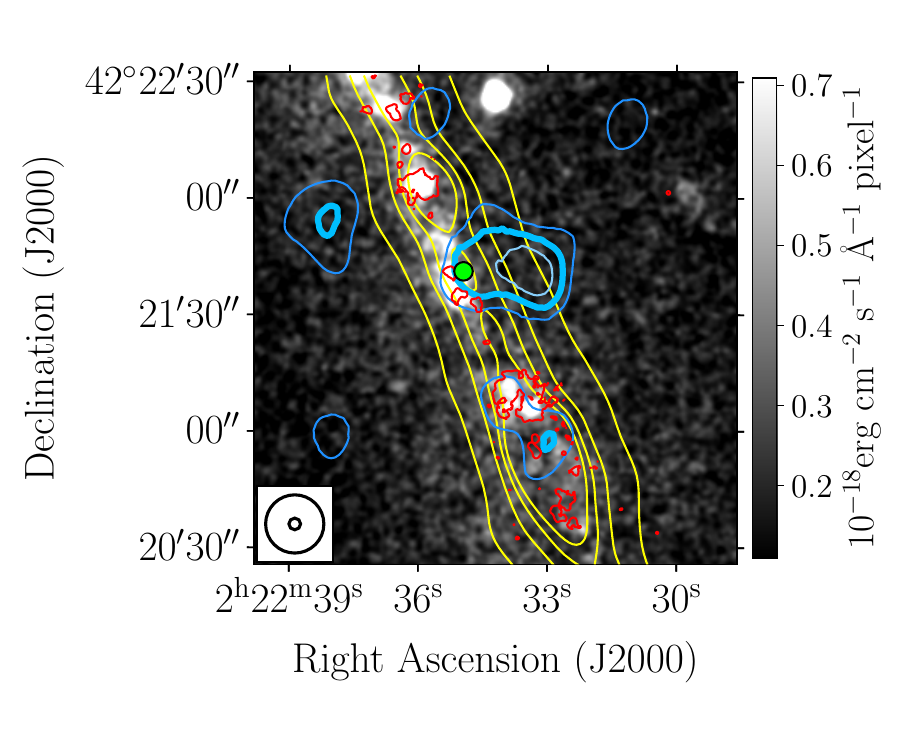}
        \put(28,67){\color{black}\setlength\fboxsep{1pt}\fcolorbox{black}{white}{\textbf{(c)}}}
    \end{overpic}

    \caption{a: Faraday Dispersion Function (FDF) spectrum at the location with $RA = 02^{\rm h}22^{\rm m}34^{\rm s}.954$ and $DEC = +42\arcdeg 21\arcmin 41\farcs20$ (indicated by a green dot in panels b and c), showing three peaks above the detection threshold. b: Far-UV image of the same region overlaid with yellow Stokes $I$ contours, red H$\alpha$ contours, and blue polarized intensity contours at $6 \sigma$, $8 \sigma$, and $10 \sigma$, corresponding to the Faraday depth of $-630\ \rm rad\ m^{-2}$.
    The $6\sigma$ contour is shown in the darkest blue, the $8\sigma$ contour is the thickest and shown in an intermediate blue, and the $10\sigma$ contour is the lightest blue.
    c: Same as b, but for the Faraday depth of $+365\ \rm rad\ m^{-2}$. The circles in the lower-left corner indicate the beam sizes: the larger circle shows the radio beam, and the smaller circle shows the far-UV pointspread function.}
    \label{fig:ngc891_fdf_spectrum_3peaks_UVIT}
\end{figure}

The S-band data were imaged with multi-scale tclean \citep{cornwell2008multiscale} and setting the gridder to mosaic as there are two pointings for NGC~891 in each observation. Figure~\ref{fig:Stokes_I_images} shows our high angular resolution (FWHM = $5\arcsec$), intermediate resolution (FWHM = $10\arcsec$), and low resolution (FWHM = $20\arcsec$) primary beam corrected total intensity images of NGC~891. These different resolutions were obtained by applying uv-tapering; the corresponding imaging parameters are listed in Table~\ref{tab:3maps}.

\setlength{\tabcolsep}{6pt}
\begin{deluxetable*}{lccc}
\tabletypesize{\normalsize}
\tablewidth{0pt} 
\tablecaption{NGC 891 Stokes $I$ maps parameters \label{tab:3maps}}
\tablehead{
\multicolumn{1}{l}{Parameters} & \colhead{} & \colhead{Stokes $I$ images} & \colhead{} \\
} 
\renewcommand{\arraystretch}{1.5} 
\startdata 
Figure label & Figure 1(a) & Figure 1(b) &  Figure 1(c) \\ 
uv weighting & Briggs Rob = 0 & Briggs Rob = 0 & Briggs Rob = 0 \\
uvtaper ($\arcsec$)$^a$ & None & 10 & 20 \\
$b_{\rm maj},b_{\rm min},PA^b$ & $4\farcs92, 4\farcs65, 1\fdg20$ & $10\farcs10, 9\farcs83, -25\fdg93$ & $20\farcs65, 18\farcs83, -30\fdg94$ \\
noise$^c$ ($\rm \mu Jy \hspace{1mm} Beam^{-1}$) & 3.4 & 5.4 & 11.5 \\
\enddata
\tablecomments{These parameters correspond to the images represented in Figure~\ref{fig:Stokes_I_images}. \\  
$^a$ The uvtaper was entered in casa software as the FWHM size of the desired beam after tapering.\\
$^b$ Synthesized beam major and minor axis and position angle. \\ $^c$ rms map noise at maximum sensitivity. The primary beam attenuation is $\lesssim 0.95$ in the C-band mosaic at the edge of NGC~891.}
\end{deluxetable*}

\subsection{RM synthesis} \label{sec: RM synthesis}
Due to the birefringence of the magneto-ionic medium, the polarization angle of a linearly polarized wave rotates as a function of wavelength. This effect is called Faraday rotation. By defining $\phi$ as the Faraday depth, the amount of rotation is $\phi \lambda^2$, and 
\begin{equation}
\phi = 0.81 \int_{z}^{0} n_e\ B_{||}dl,
\end{equation}
where $n_e$ is the thermal electron density in $\rm cm^{-3}$, and $B_{||}$ is the strength of the magnetic field component along the line of sight in $\rm \mu Gauss$ and $dl$ is an infinitesimal path length in parsecs. The integral is along the line of sight, from the position of the polarized emission ($z<0$) to the observer ($z=0$). When the magnetic field is directed toward the observer, $\phi > 0$.  

The polarized signal is comparatively weak when measured against the total intensity. To reduce noise, it is beneficial to average over a broad frequency range. However, averaging over data affected by Faraday rotation results in the signal diminishing to zero. The solution to this problem is to first reverse the effects of Faraday rotation before averaging the data. This essentially involves applying a Fourier transform. This technique is known as Faraday Rotation Measure Synthesis (RM synthesis) \citep{burn1966depolarization, brentjens2005faraday}. RM Synthesis allows us to separate polarized emission coming from regions of different Faraday depths. All RM synthesis presented in this paper applies the RM-Tools package \citep{purcell2020rm, vaneck2026}. 

The Faraday dispersion function (FDF), also known as the Faraday depth spectrum, is defined as 
\begin{equation}
\tilde{F}(\phi) = \frac{\int_{-\infty}^{+\infty} W(\lambda^2) P(\lambda^2) e^{-2i\phi\lambda^2}d\lambda^2}{\int_{-\infty}^{+\infty} W(\lambda^2)d\lambda^2},
\end{equation}
where $\rm W(\lambda^2)$ is a weighting function that is positive for any $\rm \lambda$ for which we have data and zero elsewhere, and $\rm P(\lambda^2)$ is the complex polarization of the source, defined as $\rm P(\lambda^2) = Q(\lambda^2) + i U(\lambda^2)$.

The Rotation Measure Transfer Function (RMTF), is the Fourier transform of the weighting function,
\begin{equation}
RMTF = \frac{\int_{-\infty}^{+\infty}W(\lambda^2)e^{-2i\phi \lambda^2}d\lambda^2}{\int_{-\infty}^{+\infty}W(\lambda^2)d\lambda^2}
\end{equation}
The RMTF in RM synthesis is similar to the point spread function in imaging. 
By defining $\rm F(\phi)$ as the Fourier transform of the polarization of the source, it turns out that applying RM synthesis results in a complex convolution,
\begin{equation}
\tilde{F}(\phi) = F(\phi) \ast RMTF (\phi)
\end{equation}
We used the RM CLEAN algorithm \citep{heald2008faraday} for deconvolution. 

To get a larger frequency coverage for doing RM synthesis, we concatenated the Stokes Q, U, and I cubes in S-band, and C-band. The galaxy was observed in C-band with VLA in the D and C configurations \citep{irwin2012continuum, wiegert2015chang}.

Using our concatenated S-band + C-band data with $\lambda^2$ ranging from $1.8\ \times 10^{-3}\ \rm m^2$ to $0.023\ \rm m^2$, ($\Delta \lambda^2= 2.09\ \times 10^{-2}\ \rm m^2$), and a channel spacing of $\delta \lambda^2 = 1.71\ \times 10^{-4}\ \rm m^2$ (at $3\ \rm GHz$ for frequency spacing of $28\ \rm MHz$), we obtain a theoretical Faraday‐depth resolution of $\delta \phi = 165\ \rm rad\ m^{-2}$, a maximum observable Faraday depth $\phi_{max} = 10123\ \rm rad\ m^{-2}$, and sensitivity to Faraday structures up to a scale of $\rm \Delta \phi_{max-scale} = 1720\ \rm rad\ m^{-2}$.
Compared to S-band alone ($\delta \phi = 200\ \rm rad\ m^{-2}$, $\rm \Delta \phi_{max-scale} = 560\ \rm rad\ m^{-2}$), the addition of C-band broadens the $\lambda^2$ coverage, leading to improved Faraday depth resolution and enhanced sensitivity to larger-scale Faraday structures in $\phi$-space.
For the RM synthesis, we used a Faraday depth range of $-4000 \leq \phi \leq 4000\ \rm rad\ m^{-2}$ with a sampling interval of $5\ \rm rad\ m^{-2}$.

The Stokes Q, U, and I cubes are the inputs for the RM synthesis code \citep{heald2009westerbork} and are prepared as follows: S-band measurement sets are divided into 16 spectral windows corresponding to a frequency spacing of $128\ \rm MHz$ and $\rm \lambda^2$ spacings between $3.69\ \rm cm^2$ and $26.34\ \rm cm^2$. We divided each spectral window into four quarters, each containing 14 channels, assuming that the edge channels are always flagged. This corresponds to a frequency spacing of $28\ \rm MHz$ or $\lambda^2$ spacings between $0.73\ \rm cm^2$ and $5.84\ \rm cm^2$. 
To obtain a similar range of $\lambda^2$ spacing in C-band, we used the default spectral windows in C-band without further division, ensuring consistency for RM synthesis. This corresponds to a frequency spacing of $128\ \rm MHz$ and $\lambda^2$ spacings between $0.67\ \rm cm^2$ and $1.77\ \rm cm^2$.

The tclean task was run on Stokes Q, U and I of each of these frequency intervals using these settings: A pixel size of $1\arcsec$, an image size of $2048 \times 2048$ pixels, and multi-frequency synthesis (MFS) with nterms set to $2$. A maximum of $100,000$ clean components was applied to reach the clean threshold of $0.01\ \rm mJy$.
In all cases, the tclean reached the specified clean threshold before the iteration limit was reached.
Briggs weighting with a robust parameter of $0$ was used, along with a gridder set to widefield and a uvtaper resulting in a beam size of $10\arcsec$. The result images were convolved to a common angular resolution of $15\arcsec$ using the imsmooth task in CASA. 

We also constructed a synchrotron-only Stokes I cube. We used the $15\arcsec$ resolution thermal fraction map of NGC~891 at $6\ \rm GHz$ from the CHANG-ES thermal emission analysis in C-band and L-band by \citep[][Irwin et al. in prep]{vargas2018chang}. Assuming a thermal spectral index of $\alpha=-0.1$, the thermal emission cube was extrapolated to all frequencies used in the RM synthesis.

\subsection{Ancillary data}
In order to facilitate interpretation of the radio polarization data, several archival observations of NGC~891 were considered for analysis. We focus here on image data that provide insight in distribution of Faraday rotating plasma in NGC~891, which is mainly the warm ionized medium \citep{heiles2012magnetic}, or the young massive stellar population that ionizes this medium. For an edge-on galaxy such as NGC~891, tracers that are particularly sensitive or insensitive to interstellar extinction, or data with velocity information are useful tracers of the 3-dimensional distribution of the interstellar medium.

\subsubsection{X-rays}

NGC~891 was observed by ROSAT \citep{bregman1994x}, 
XMM-Newton \citep{temple2005x, hodges2018hot}, 
and Chandra \citep{strickland2004high, temple2005x, LiWang2013a, LiWang2013b, hodges2018hot}. 
These observations show diffuse X-ray emission around the center of NGC~891, and a secondary region in the north-west side, around 
$(RA ,DEC) =(02^{\rm h}22^{\rm m}36^s, +42\arcdeg22\arcmin22\arcsec)$.
The XMM data suffered from a flare of an ultraluminous X-ray binary in NGC~891. 
This paper employs two Chandra datasets, previously published by \citet{temple2005x} and \citet{hodges2012deep}, contained in the Chandra Data Collection ~\dataset[DOI:10.25574/cdc.601]{https://doi.org/10.25574/cdc.601}. 
We subtracted the compact sources to obtain a $0.3 - 8\ \rm keV$ image of the diffuse emission, similar to that presented by \citet{hodges2012deep}. The resulting diffuse X-ray map is shown in Figure~\ref{fig:Xray_FDF}.

\subsubsection{\texorpdfstring{H$\alpha$ intensity}{H alpha intensity}}
$\rm H\alpha$ intensity is a useful tracer of the denser WIM. $\rm H \alpha$ brightness is proportional to emission measure, which scales with the square of the electron density, unlike Faraday rotation, which scales linearly with the electron density. Significant Faraday rotation may arise in a region with emission measure that is far below what is detectable \citep{uyaniker2003radio}. Sometimes, Faraday rotation of a few hundred to a few thousand $\rm rad\ m^{-2}$ is associated with bright H\,{\footnotesize II} regions \citep{costa2016denser,costa2018faraday}.

In this paper we use a stellar continuum subtracted $\rm H\alpha$ image previously published by \citet{vargas2018chang}. Table~\ref{tab:HIIregions} lists H\,{\footnotesize II} regions and complexes considered in the analysis. We also reproduced a Hubble Space Telescope (HST) image using data from the HST Legacy Archive, accessible directly as ~\dataset[DOI: 10.17909/7r5b-cf06]{https://doi.org/10.17909/7r5b-cf06}, first published and analyzed in detail by \citet{rossa2004hubble}.

Kinematic information for the WIM is derived from the $\rm H\alpha$ velocity cube of \citet{kamphuis2007} made with the TAURUS II imaging Fabry-Pérot spectrograph at the William-Herschel Telescope on La Palma. The angular resolution is limited by seeing of approximately $2\arcsec$. The velocity resolution is $40.7\ \rm km\ s^{-1}$ (FWHM). The $\rm H\alpha$ cube covers the inner $10\arcmin$ of NGC~891 in two slightly overlapping fields. \citet{kamphuis2007} derived an extinction optical depth toward the center of NGC~891 of $\tau_{H\alpha} \sim 6$. The observable H\,{\footnotesize II} regions are necessarily only those that suffer modest extinction. Spatially resolved kinematics of the extraplanar diffuse ionized gas in NGC~891 was done by \citet{Boettcher2016} and \citet{Lu2024}.

\begin{figure}
    \centering

    \begin{overpic}[trim=90 0 100 80, clip, width=0.95\columnwidth]{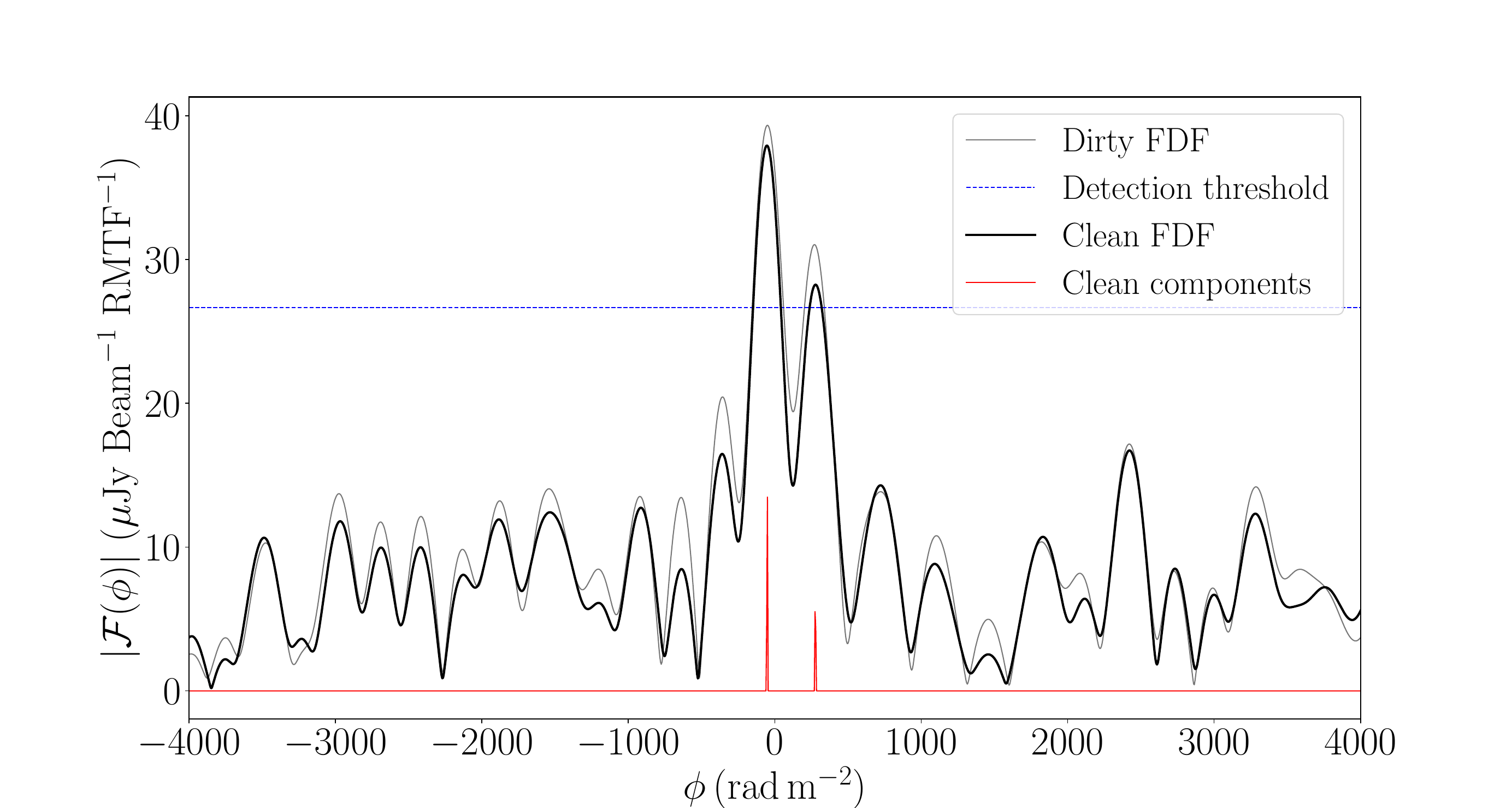}
        \put(10,50){\color{black}\setlength\fboxsep{1pt}\fcolorbox{black}{white}{\textbf{(a)}}}
    \end{overpic}

    \begin{overpic}[trim=10 30 5 20, clip, width=0.95\columnwidth]{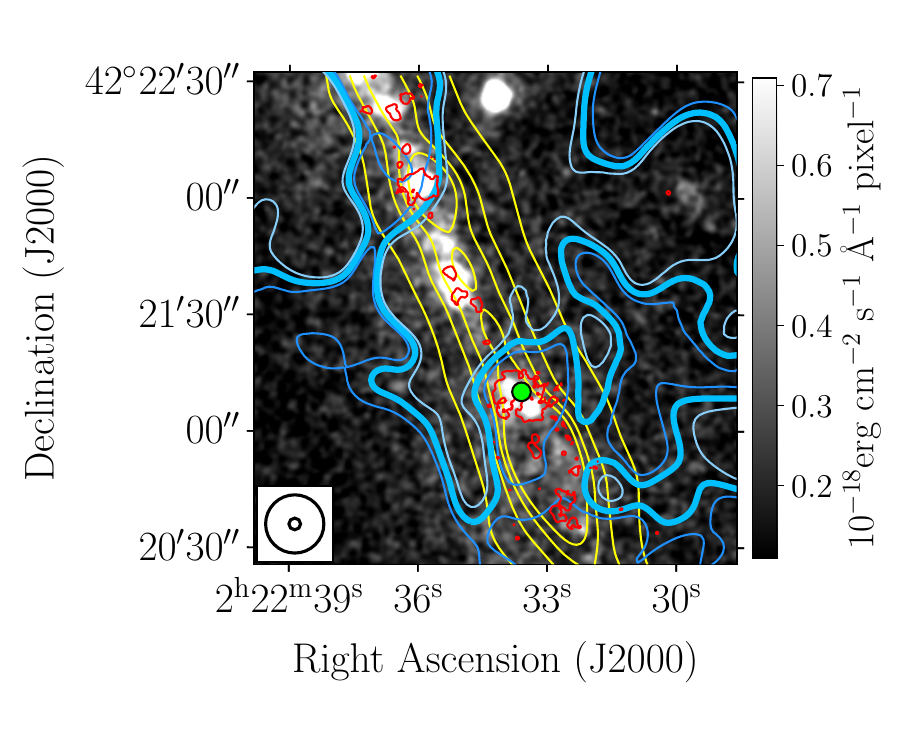}
        \put(28,66){\color{black}\setlength\fboxsep{1pt}\fcolorbox{black}{white}{\textbf{(b)}}}
    \end{overpic}

    \begin{overpic}[trim=10 30 5 20, clip, width=0.95\columnwidth]{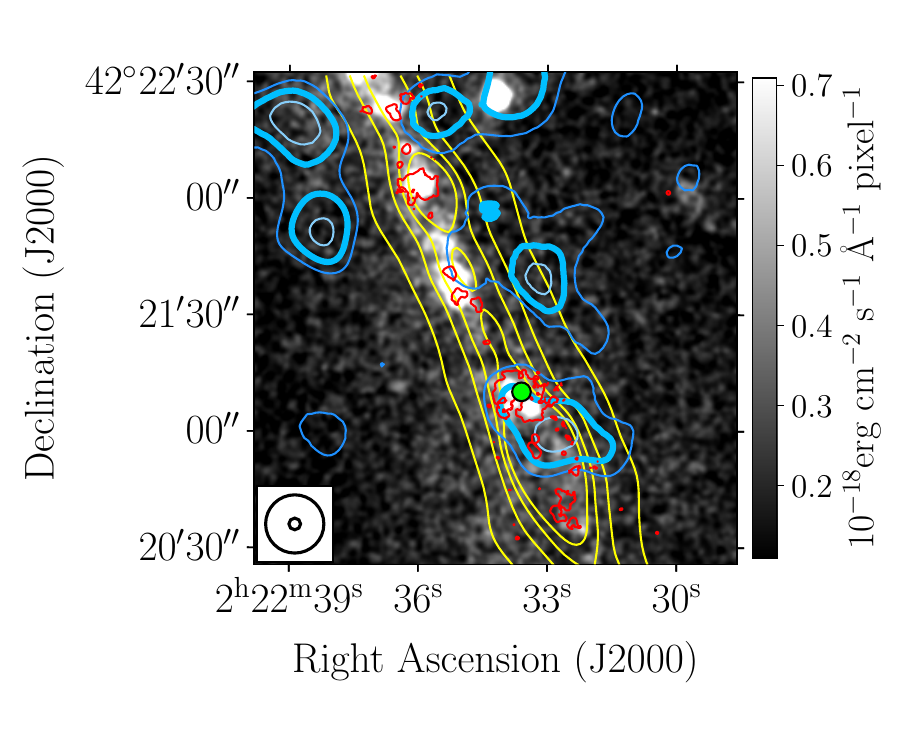}
        \put(28,66){\color{black}\setlength\fboxsep{1pt}\fcolorbox{black}{white}{\textbf{(c)}}}
    \end{overpic}

    \caption{a: FDF spectrum at the location with $RA = 02^{\rm h}22^{\rm m}33\fs602$ and $DEC = +42\arcdeg 21\arcmin 10\farcs20$, showing two peaks above the $8 \sigma$ detection threshold. b: Far-UV image of the same region overlaid with yellow Stokes $I$ contours, red H$\alpha$ contours, and blue polarized intensity contours at $6 \sigma$, $8 \sigma$, and $10 \sigma$, corresponding to the Faraday depth of $+120\ \rm rad\ m^{-2}$. At this Faraday depth, the target location (indicated by a green dot) shows no detectable polarized emission, while the surrounding regions are bright. 
    The $6\sigma$ contour is shown in the darkest blue, the $8\sigma$ contour is the thickest and shown in an intermediate blue, and the $10\sigma$ contour is the lightest blue.
    c: Same as b, but for a Faraday depth of $+280\ \rm rad\ m^{-2}$, where the target location itself exhibits significant polarized emission. The circles in the lower-left corner indicate the beam sizes: the larger circle shows the radio beam, and the smaller circle shows the far-UV pointspread function.}
    \label{fig:ngc891_fdf_spectrum_2peaks_UVIT}
\end{figure}

\subsection{Ultraviolet}
The ultraviolet (UV) is an interesting wavelength range because interstellar extinction is much stronger in the UV. Association of H\,{\footnotesize II} regions with star clusters in the UV reinforces the notion that these H\,{\footnotesize II} regions are located on the near side of NGC 891. \citet{seon2014} pointed out that the diffuse ultraviolet is starlight scattered by dust in the halo. 

NGC~891 was observed by the GALEX \citep{martin2005galaxy} mission in near and far UV, and by the UVIT telescope on board the ASTROSAT mission \citep{kumar2012ultraviolet, subramaniam2016orbit,joseph2025}. Both data sets were considered in the analysis, but UVIT has an angular resolution $<2\farcs5$, compared with $4\arcsec$-$6\arcsec$ for GALEX.

UVIT Level 1 data from 7 orbits on 6 December 2022, with a total on-source exposure time of 7.48 ks were processed by S. Ranasinghe using the procedures and software described in \citet{ranasinghe2025user}. Only the far-UV detector was operational at the time of observation. The filter was the F148W filter with central wavelength 148.1 nm and width 50 nm \citep{tandon2020}. To reduce noise from photon counting statistics, the image was convolved with a Gaussian of 4 pixels (FWHM), for a final angular resolution of $3\farcs0$. The astrometry was corrected by registering the co-added UVIT image against the GALEX far-UV image, with r.m.s. residuals of $0\farcs3$. 

\section{Results} \label{sec:Results}
\subsection{Total intensity images}
The Stokes $I$ images of NGC~891 at three different angular resolutions are shown in Figure~\ref{fig:Stokes_I_images}.  Figure~\ref{fig:Stokes_I_images}a presents the total intensity map at the highest available resolution of $4\farcs 92 \times 4\farcs65$, which reveals the fine structure along the galaxy. Figure~\ref{fig:Stokes_I_images}b shows the intermediate-resolution image at $10\farcs10 \times 9\farcs83$, and Figure~\ref{fig:Stokes_I_images}c displays the low-resolution image at $20\farcs65 \times 18\farcs83$. The corresponding image characteristics are summarized in Table~\ref{tab:3maps}. The rms noise level is $3.4\ \rm \mu Jy\ Beam^{-1}$ in the high-resolution map and increases to $5.4\ \rm \mu Jy\ Beam^{-1}$ in the intermediate-resolution image and to $12.5\ \rm \mu Jy\ Beam^{-1}$ in the low-resolution image due to the larger synthesized beam. The high-resolution image clearly shows the thin disk of NGC~891 and allows identification of compact emission features along the midplane. In contrast, the lower-resolution maps are more sensitive to diffuse halo emission, extending several kpc above and below the disk plane. The $5\arcsec$ resolution image was used by \citet{heesen2025}, who derived a scale height $0.13 \pm 0.01$ kpc for the thin disk and $1.00 \pm 0.09$ kpc for the thick disk.

The 3 GHz flux density of NGC~891 was measured to be $366 \pm 1\ \rm mJy$ from the $20\arcsec$ image corrected for primary beam attenuation. The stated error includes only the effect of noise within the aperture over which the intensity was integrated.
\citet{schmidt2019chang} presented a detailed analysis for the total flux of NGC~891 at 1.5 GHz and at 6 GHz, including Effelsberg observations at 4.85 GHz, scaled to 6 GHz.
To estimate the fraction of the total flux recovered in our VLA observations we interpolate the 1.5 GHz flux density $737\pm37\ \rm mJy$ and the 6 GHz flux density $252\pm27\ \rm mJy$, assuming a powerlaw spectrum, to 3 GHz for a total flux density $431\pm33\ \rm mJy$. This may be compared with the flux density of $430 \pm 60\ \rm mJy$ at $2.695\ \rm GHz$ by \citet{kazes1970} \citep[see also][]{hummel1991}. This suggests our VLA observations recovered $85\%$ of the total flux density at 3 GHz. The missing flux density is twice the uncertainty in the total flux density. This does not include the uncertainty in the absolute flux calibration of our observations which is approximately $10\%$. 

\setlength{\tabcolsep}{1mm}
\begin{deluxetable}{ccccc}
\tabletypesize{\scriptsize}
\tablewidth{0pt} 
\tablecaption{H\,{\footnotesize II} regions in the disk of NGC 891 with their positions and peak brightness. \label{tab:HIIregions}}
\tablehead{
\colhead{Label} & 
\colhead{\begin{tabular}{c} RA \\ (h m s) \end{tabular}} & 
\colhead{\begin{tabular}{c} DEC \\ ($\arcdeg\ \arcmin\ \arcsec$) \end{tabular}} & 
\colhead{\begin{tabular}{c} Peak Brightness \\ (intensity\ unit$^a$) \end{tabular}} &
\colhead{Cross ID$^b$}
} 
\renewcommand{\arraystretch}{1.5} 
\startdata
$1$  & $02\ 22\ 24.784 $ & $42\ 17\ 18.97$ & $1.942$ & $-$\\
$2$  & $02\ 22\ 25.183$ & $42\ 17\ 43.10$ & $6.174$ & $-$\\
$3$  & $02\ 22\ 27.490$ & $42\ 18\ 24.90$ & $6.223$ & $-$\\
$4$  & $02\ 22\ 28.840$ & $42\ 18\ 54.19$ & $4.551$ & $-$\\
$5$  & $02\ 22\ 29.139$ & $42\ 19\ 12.02$ & $2.183$ & $-$\\
$6^c$  & $02\ 22\ 30.500$ & $42\ 19\ 53.92$ & $3.012$ & $-$\\
$7$  & $02\ 22\ 32.431$ & $42\ 20\ 36.81$ & $2.206$ & $-$\\
$8$  & $02\ 22\ 32.530$ & $42\ 20\ 42.05$ & $2.393$ & $-$\\
$9$  & $02\ 22\ 32.812$ & $42\ 20\ 38.87$ & $2.581$ & $-$\\
$10$ & $02\ 22\ 33.403$ & $42\ 21\ 5.08$ & $12.28$ & $\rm WF3-1$\\
$11$ & $02\ 22\ 33.882$ & $42\ 21\ 11.33$ & $15.30$ & $\rm WF3-1$\\
$12$ & $02\ 22\ 34.659$ & $42\ 21\ 33.32$ & $2.980$ & $\rm WF3-2$\\
$13$ & $02\ 22\ 35.040$ & $42\ 21\ 35.38$ & $3.196$ & $\rm WF3-2$\\
$14$ & $02\ 22\ 35.234$ & $42\ 21\ 40.61$ & $4.905$ & $\rm WF3-2$\\
$15$ & $02\ 22\ 36.013$ & $42\ 22\ 4.70$ & $8.606$ & $WF3-3-SS2$\\
$16$ & $02\ 22\ 36.305$ & $42\ 22\ 13.08$ & $5.914$ & $-$\\
$17$ & $02\ 22\ 36.597$ & $42\ 22\ 22.50$ & $3.196$ & $\rm WF4-2$\\
$18$ & $02\ 22\ 37.279$ & $42\ 22\ 43.45$ & $5.594$ & $\rm WF4-SS-1$\\
$19$ & $02\ 22\ 37.285$ & $42\ 22\ 50.80$ & $4.850$ & $\rm WF4-SS-1$\\
$20$ & $02\ 22\ 37.366$ & $42\ 22\ 33.98$ & $6.069$ & $\rm WF4-SS-1$\\
$21$ & $02\ 22\ 38.061$ & $42\ 23\ 10.69$ & $1.782$ & $-$\\
$22$ & $02\ 22\ 38.148$ & $42\ 23\ 1.22$ & $4.314$ & $\rm WF4-1$\\
$23$ & $02\ 22\ 38.339$ & $42\ 23\ 3.30$ & $2.932$ & $-$\\
$24$ & $02\ 22\ 38.540$ & $42\ 23\ 15.89$ & $2.454$ & $-$\\
$25$ & $02\ 22\ 40.870$ & $42\ 24\ 17.64$ & $4.529$ & $-$\\
$26$ & $02\ 22\ 41.167$ & $42\ 24\ 31.27$ & $12.59$ & $-$\\
\enddata
\tablecomments{$^a$ intensity unit is $10^{-17}$erg\,s$^{-1}$\,cm$^{-2}$\,pixel$^{-1}$ \\ $^b$ The Cross ID column lists the corresponding H\,{\tiny II} region identifiers from \citet{rossa2004hubble} \\ $^c$ Source 6 is not associated with SN~1986J, as its position is significantly offset from the published coordinates of the supernova \citep{bietenholz2017sn}.
}
\end{deluxetable}

\subsection{Polarization maps}
\begin{figure*}
    \centering

    \begin{minipage}[t]{0.48\textwidth}
        \centering
         \begin{overpic}[trim=170 10 10 10, clip, width=\linewidth]{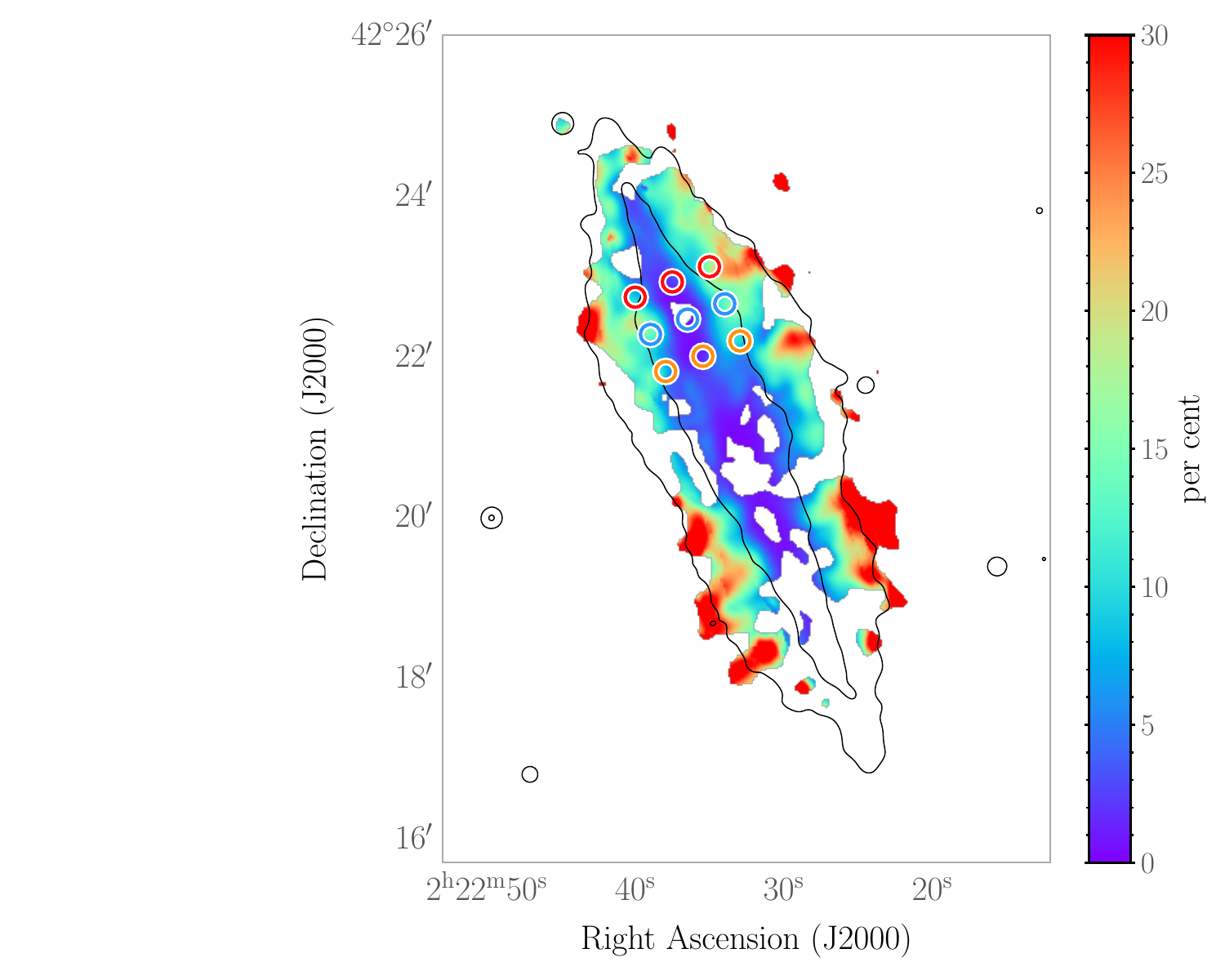}
            \put(70,90){\color{black}\setlength\fboxsep{1pt}\fcolorbox{black}{white}{\textbf{(a)}}}
        \end{overpic}
    \end{minipage}
    \hfill
    \begin{minipage}[t]{0.48\textwidth}
        \centering
        \begin{overpic}[trim=160 10 10 10, clip, width=\linewidth]{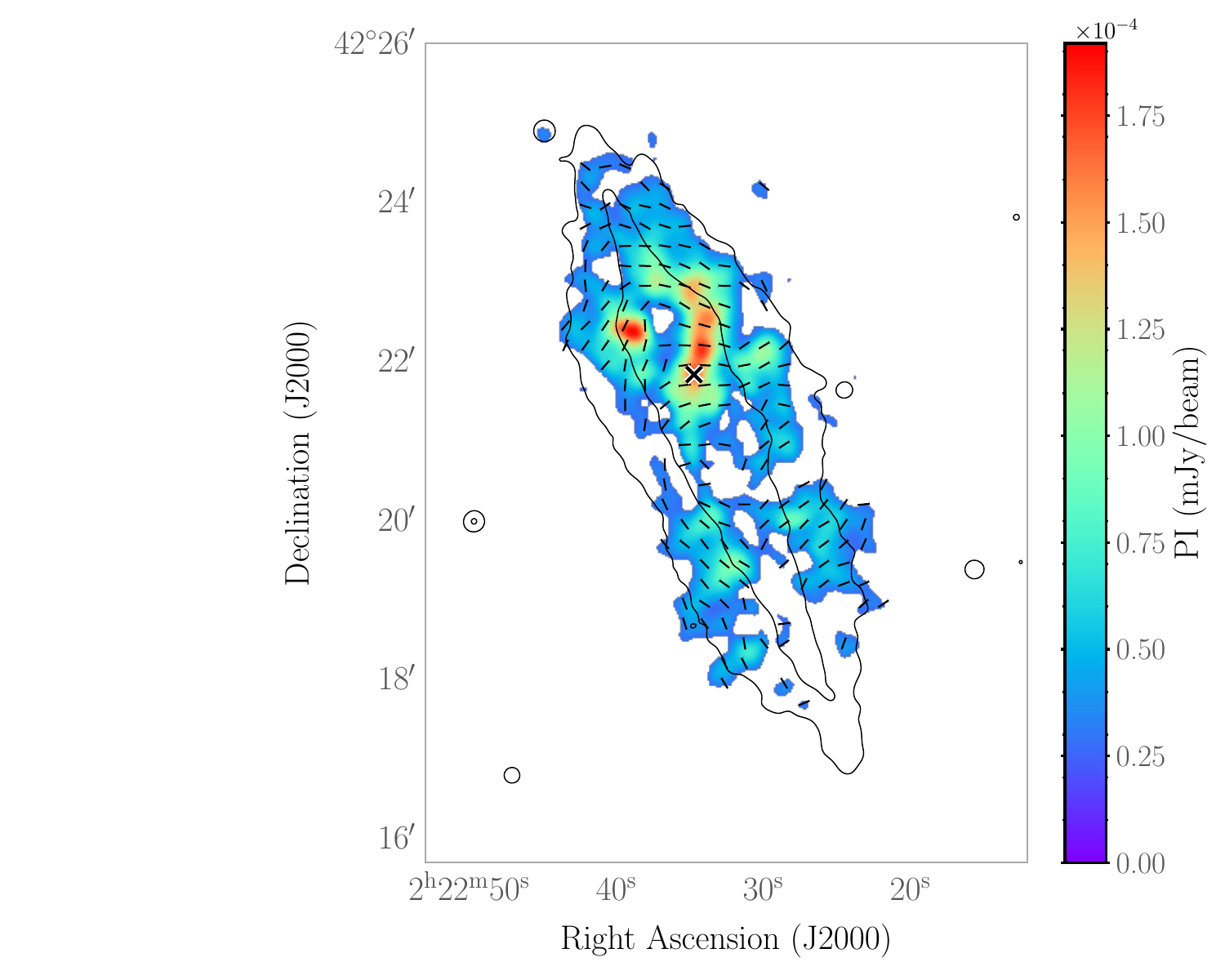}
            \put(70,89){\color{black}\setlength\fboxsep{1pt}\fcolorbox{black}{white}{\textbf{(b)}}}
        \end{overpic}
    \end{minipage}

    \vspace{2mm}

    \begin{minipage}[t]{0.48\textwidth}
        \centering
        \begin{overpic}[trim=160 10 10 10, clip, width=\linewidth]{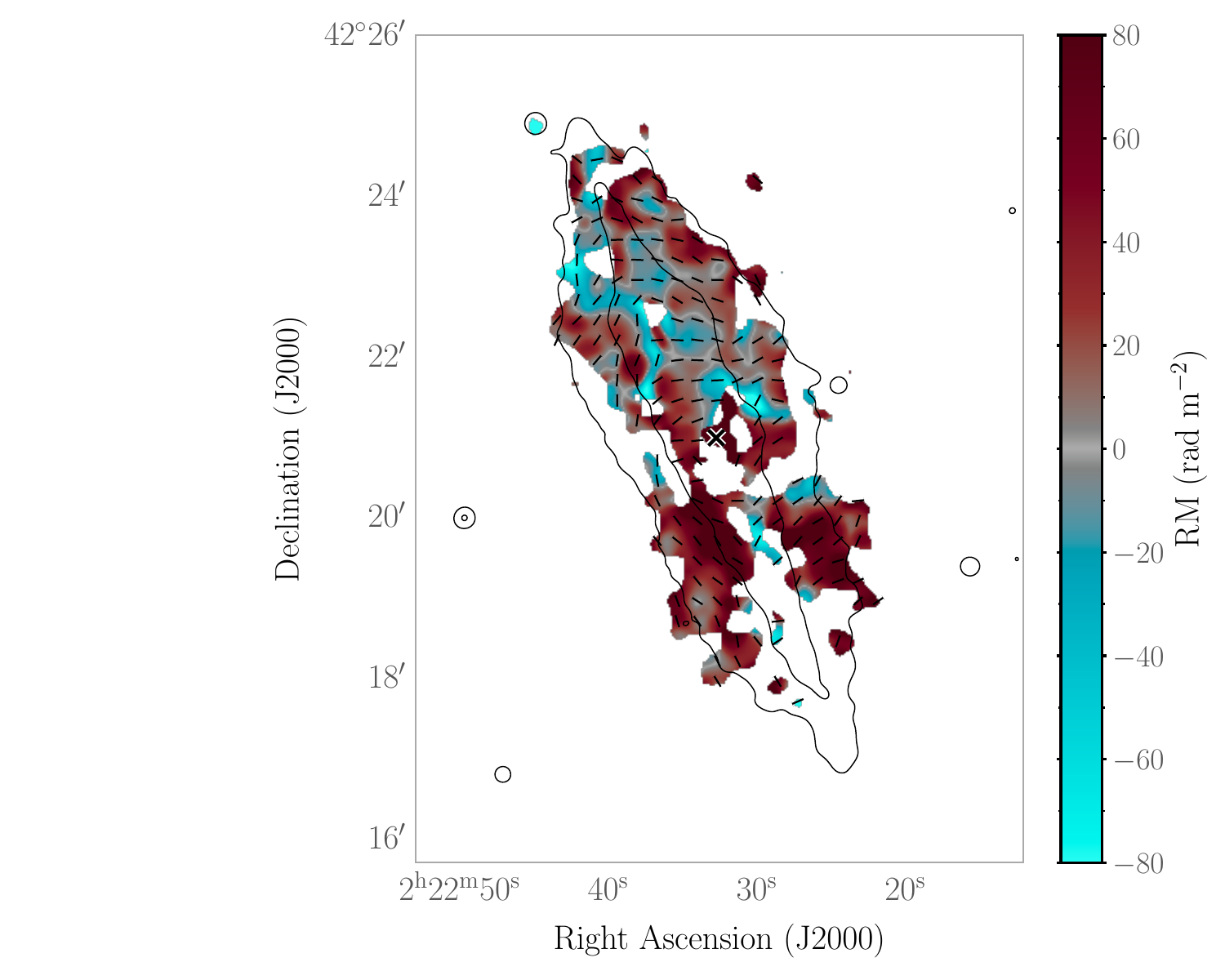}
            \put(70,90){\color{black}\setlength\fboxsep{1pt}\fcolorbox{black}{white}{\textbf{(c)}}}
        \end{overpic}
    \end{minipage}
    \hfill
    \begin{minipage}[t]{0.48\textwidth}
        \centering
        \begin{overpic}[trim=170 10 10 10, clip, width=\linewidth]{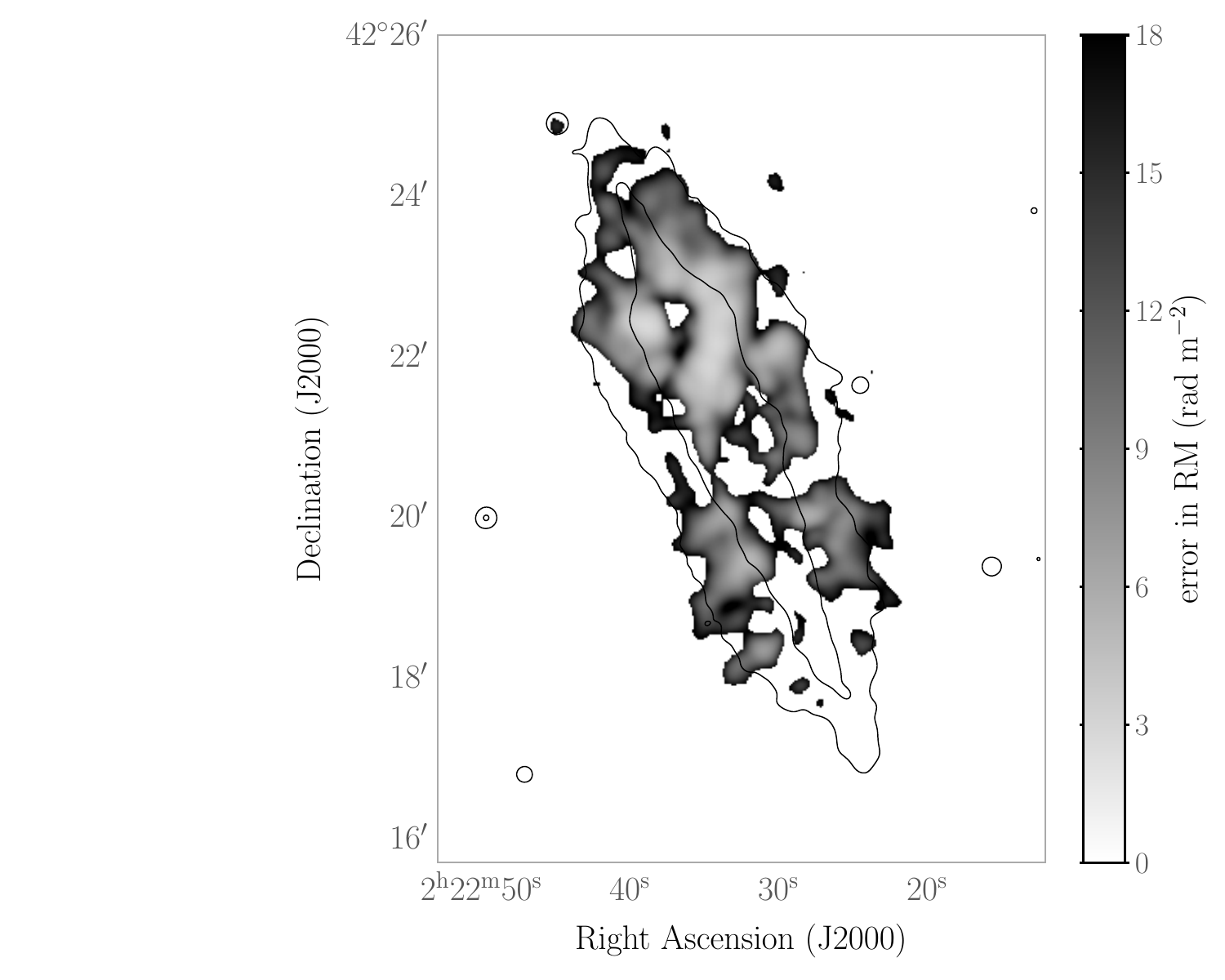}
            \put(70,90){\color{black}\setlength\fboxsep{1pt}\fcolorbox{black}{white}{\textbf{(d)}}}
        \end{overpic}
    \end{minipage}

    \caption{Polarization properties of NGC~891 from combined S-band and C-band observations with a beam size of $15 \arcsec$. The Stokes $I$ contour levels are: $18\sigma_I$ and $100\sigma_I$, with $\sigma_I = 5.4\ \rm \mu \rm Jy\ Beam^{-1}$. The vector orientations in b, and c represent polarization angle corrected for Faraday rotation. a: Percentage polarization at $3.26\ \rm GHz$, calculated using the synchrotron Stokes $I$ emission after subtraction of the thermal contribution, b: Polarized intensity at $3.26\ \rm GHz$, c: foreground-corrected Rotation Measure (foreground RM $=-55.7\ \rm rad\, m^{-2}$), d: Error in Rotation Measure (the foreground RM uncertainty of $20\ \rm rad\, m^{-2}$ is not included). The colored circles in panel a mark the nine regions selected for the depolarization analysis shown in Figure~\ref{fig:fracpol2_lambda2}, and the circle colors correspond to the colors used for the depolarization curves in that figure. The crosses in panels b and c mark the locations of Faraday depth spectra shown in Figure~\ref{fig:extreme_FDFs}.}
    \label{fig:ngc891_polarization_maps}
\end{figure*}

\begin{figure}
    \centering

    \begin{overpic}[trim=40 0 70 75, clip, width=0.95\columnwidth]{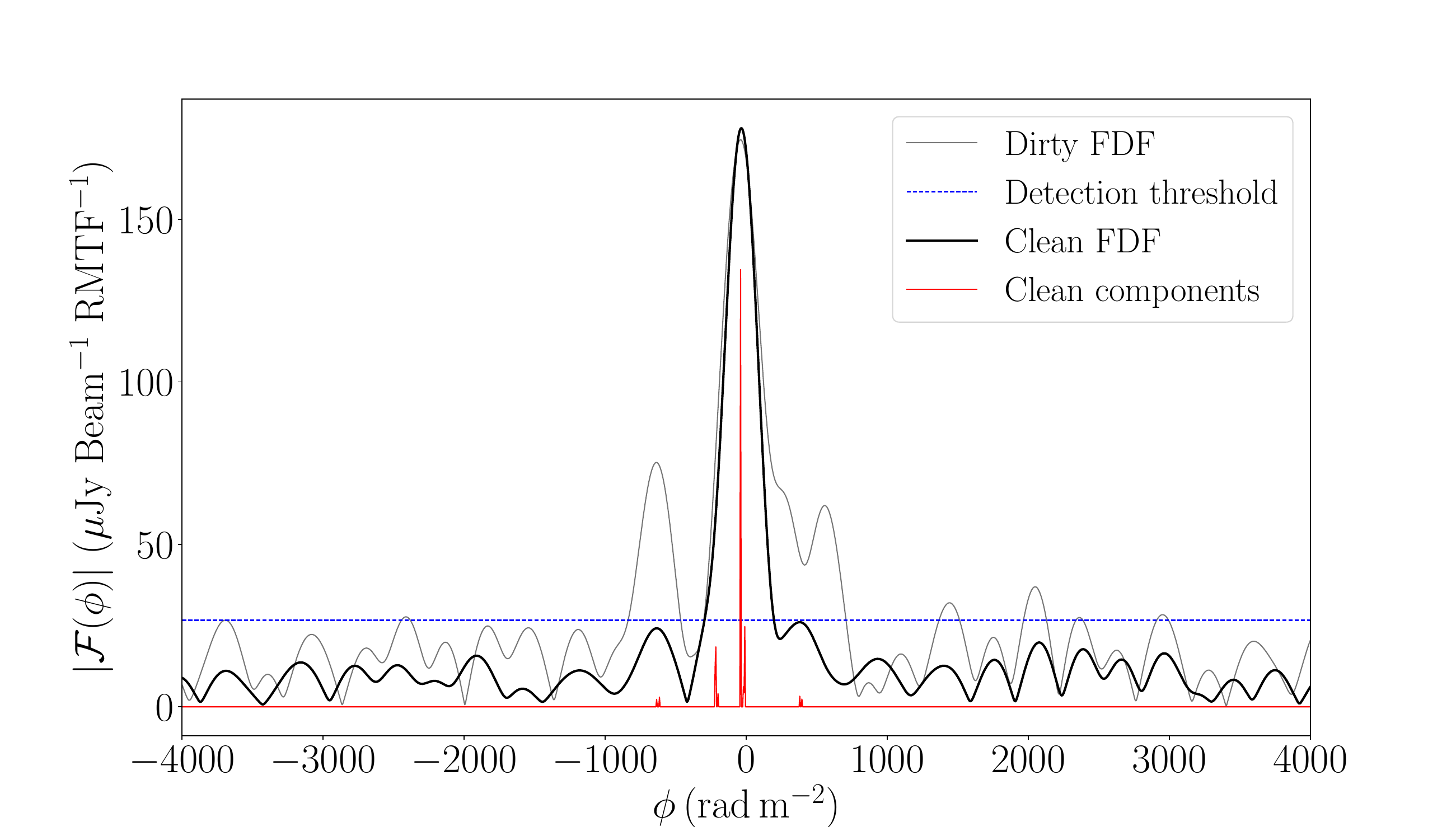}
        \put(12,50){\color{black}\setlength\fboxsep{1pt}\fcolorbox{black}{white}{\textbf{(a)}}}
    \end{overpic}

    \begin{overpic}[trim=40 0 70 75, clip, width=0.95\columnwidth]{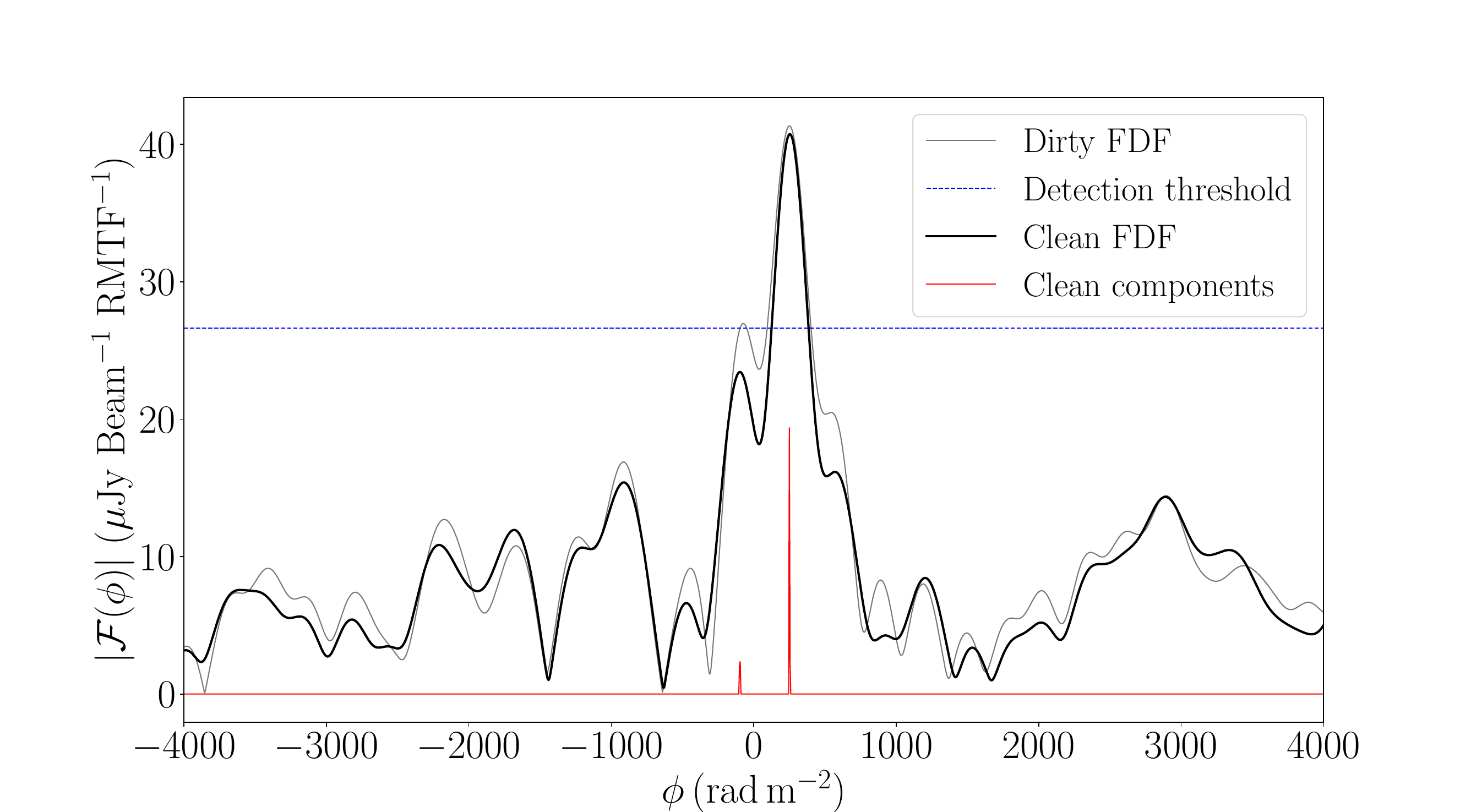}
        \put(12,48){\color{black}\setlength\fboxsep{1pt}\fcolorbox{black}{white}{\textbf{(b)}}}
    \end{overpic}

    \caption{Clean Faraday Dispersion Function (FDF) spectra at two locations in NGC~891: (a) northeastern halo region at $RA = 02^{\rm h}22^{\rm m}34\fs684,\ DEC = +42\arcdeg 21\arcmin 50\farcs20$ showing high polarized intensity, and (b) central region at $RA = 02^{\rm h}22^{\rm m}32\fs700,\ DEC = +42\arcdeg 20\arcmin 59\farcs19$ with large Faraday depth ($\phi = 250\ \rm rad\ m^{-2}$). Both exhibit pronounced peaks.}
    \label{fig:extreme_FDFs}
\end{figure}

\begin{figure*}
    \centering

    \begin{overpic}[trim=10 70 10 100, clip, width=0.95\textwidth]{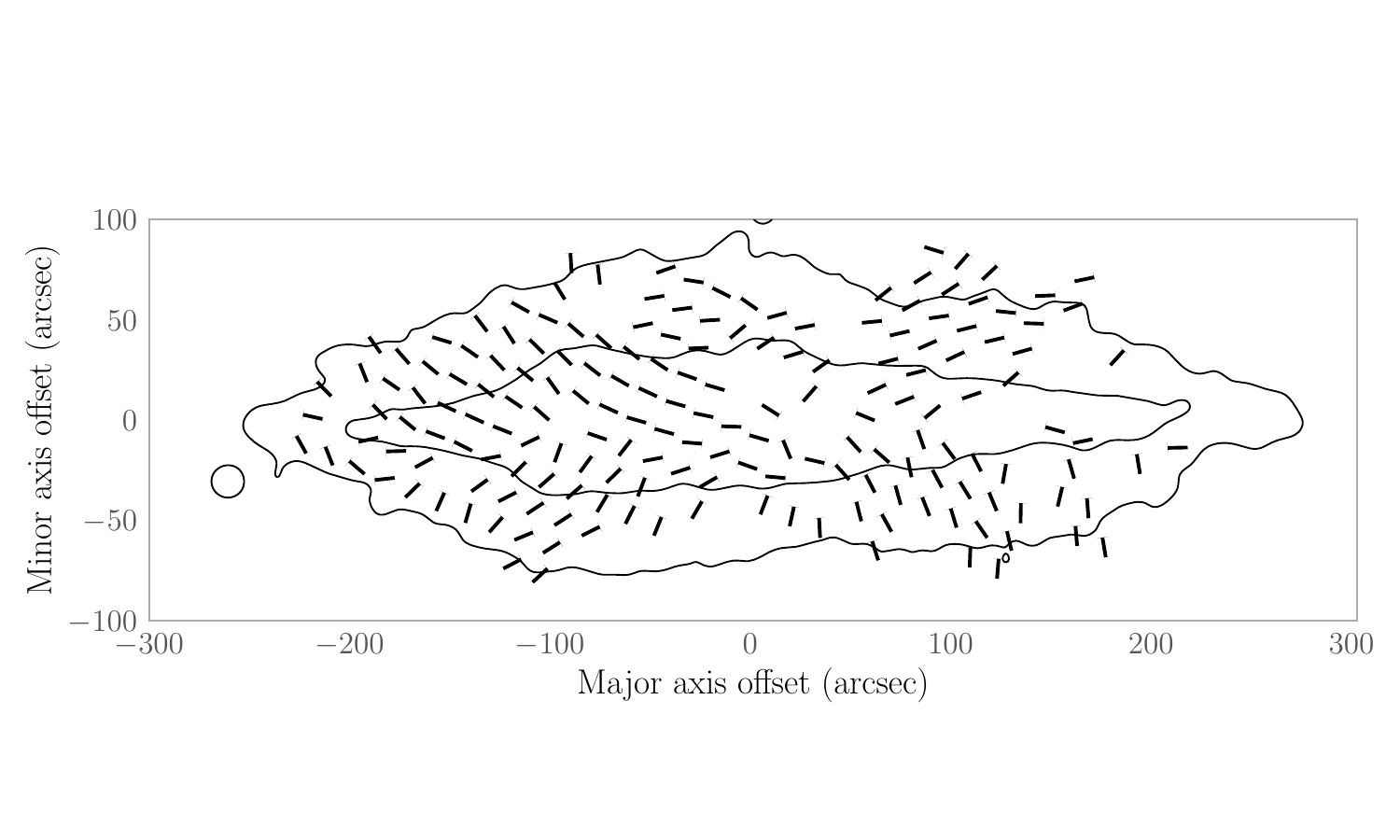}
        \put(10,32){\color{black}
        \setlength\fboxsep{1pt}
        \fcolorbox{black}{white}{\textbf{(a)}}}
    \end{overpic}

    \vspace{6mm}

    \begin{overpic}[trim=10 10 10 10, clip, width=0.95\textwidth]{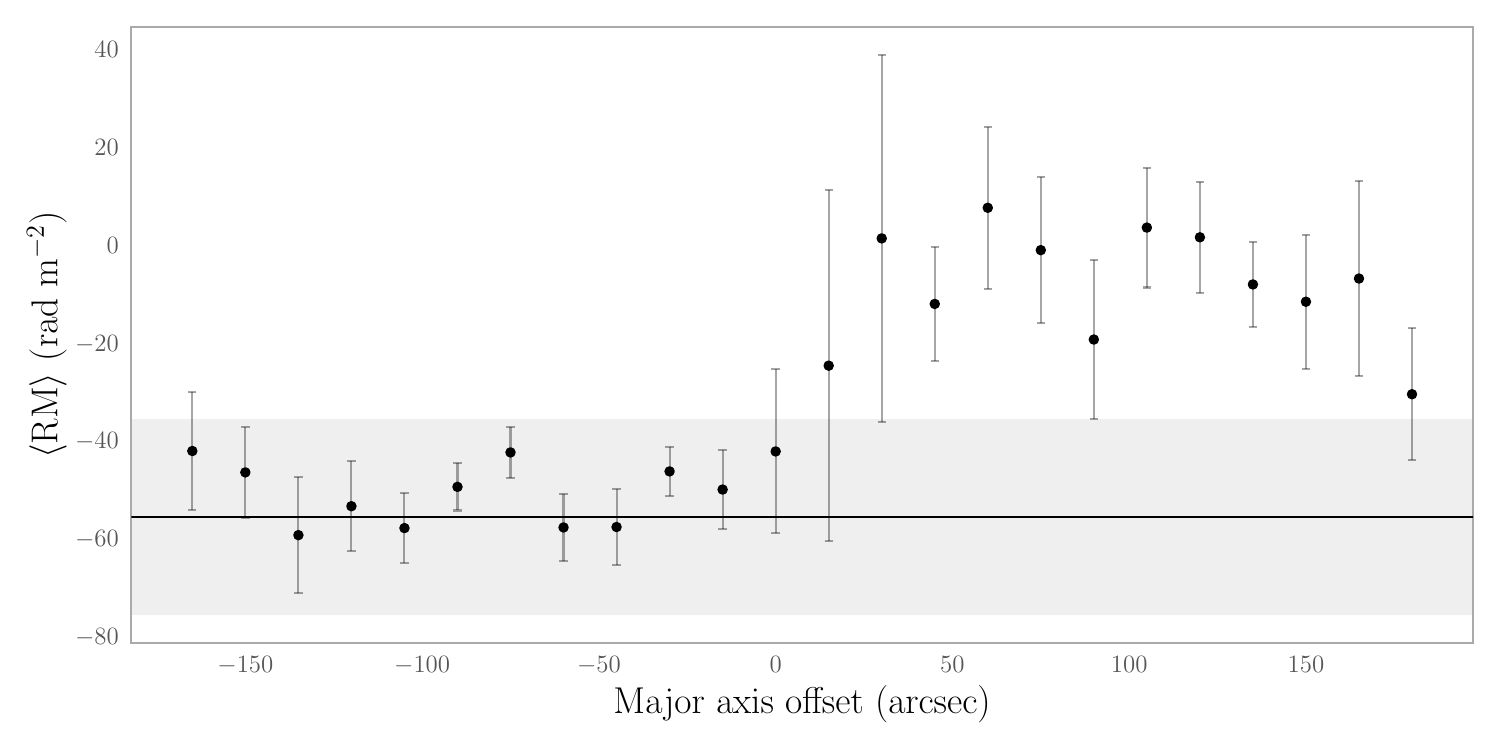}
        \put(10,44){\color{black}
        \setlength\fboxsep{1pt}
        \fcolorbox{black}{white}{\textbf{(b)}}}
    \end{overpic}

    \caption{a: Total intensity contours of NGC~891 (contour levels are: $18\sigma_I$ and $100\sigma_I$, with $\sigma_I = 5.4\ \rm \mu \rm Jy\ Beam^{-1}$) overlaid with vector orientations indicating the plane-of-sky magnetic field orientation derived from polarization angles (rotated by $90\degr$). The circle at offset $\sim 260\arcsec$ is a background radio source. b: Profile of mean Rotation Measures along the major axis, weighted by RM errors. The profile is measured along the major axis, where major axis offset of 0 is defined at RA = $02^{\rm h}22^{\rm m}33\fs37$ and DEC = $+42\arcdeg21\arcmin04\farcs69$. The horizontal line indicates the Galactic foreground rotation measure of $-55.7\ \rm rad\ m^{-2}$, as described in the text, and the gray shaded region indicates its uncertainty $\pm20\ \rm rad\ m^{-2}$. Northeast is to the left and southwest to the right in both panels.
    }

    \label{fig:AVG_RMprofileBvector}
\end{figure*}

\begin{figure*}
    \centering
    \includegraphics[trim=20 10 10 20, clip, width=\textwidth]{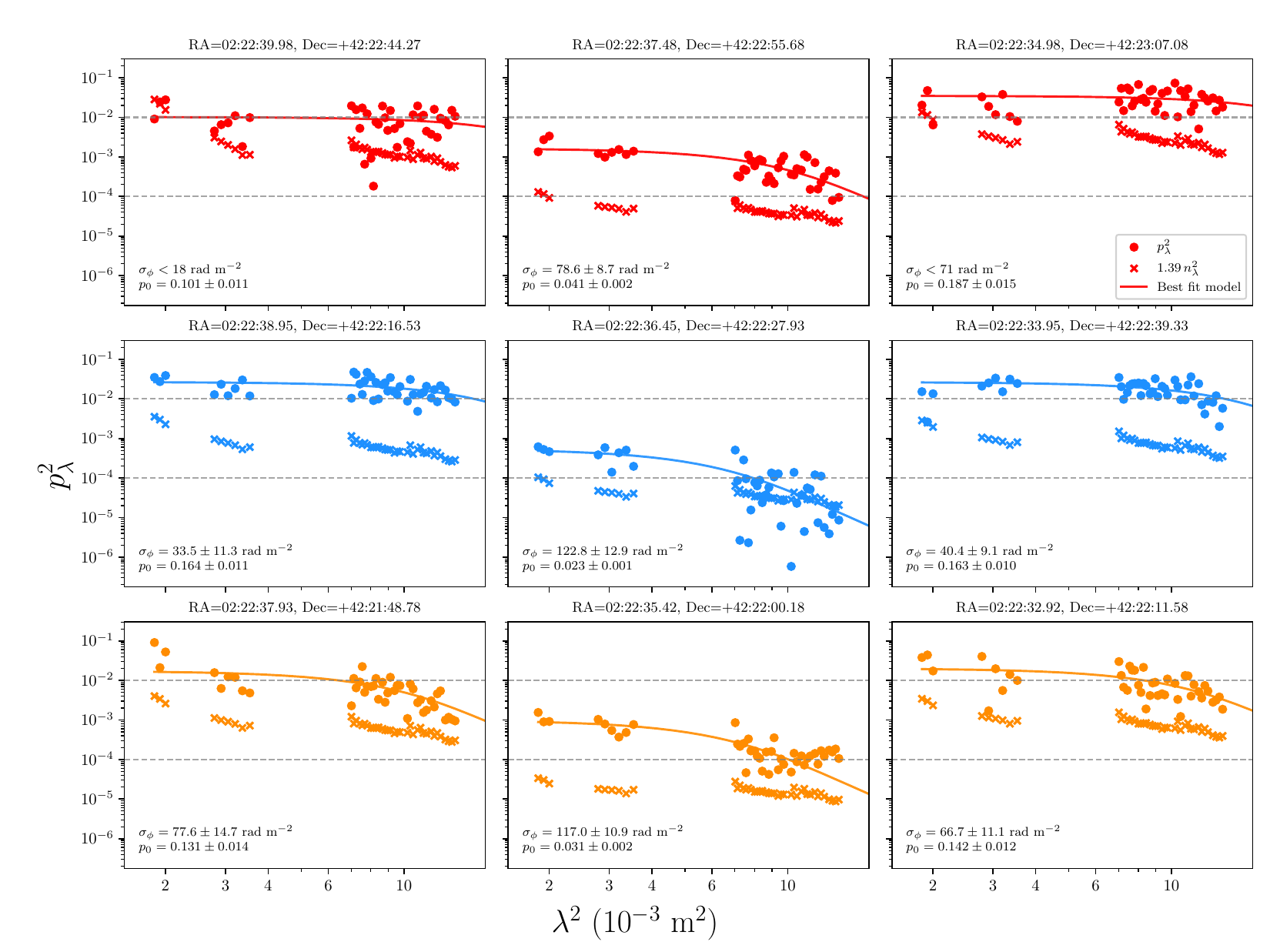}
    \caption{Fractional polarization squared ($p_\lambda^2$) versus wavelength squared ($\lambda^2$) for nine positions across NGC~891 (locations shown in Figure~\ref{fig:ngc891_polarization_maps}a). The fractional polarization was calculated using the synchrotron-only Stokes I cube after subtraction of the thermal emission contribution. 
    The crosses, $1.39\ n_\lambda^2$, indicate the expected median of $p_\lambda^2$ in the absence of polarized signal,
    where $n_\lambda$ is the noise in fractional polarization as defined in the text.
    The solid curves show the best-fit depolarization model \protect\citep{sokoloff1998depolarization}, and the best-fit parameters $p_0$ and $\sigma_\phi$ are listed in each panel. Dashed horizontal lines mark the $1\%$ and $10\%$ polarization levels.}
    \label{fig:fracpol2_lambda2}
\end{figure*}

\begin{figure*}
    \centering
    \begin{overpic}[trim=220 130 220 130, clip, width=0.5\textwidth]{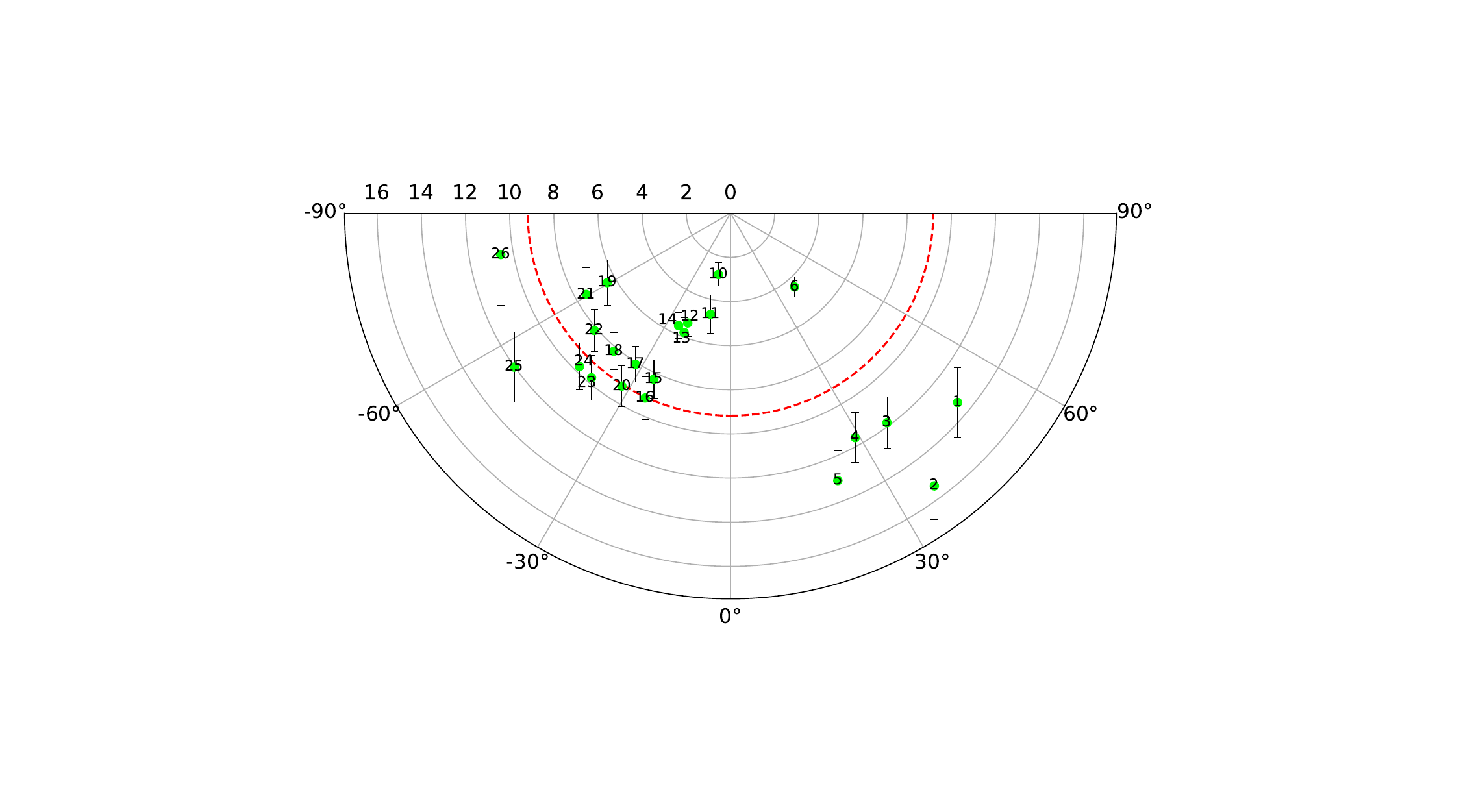}
        \put(85,44){\color{black}\setlength\fboxsep{1pt}\fcolorbox{black}{white}{\textbf{(a)}}}
    \end{overpic}

    \begin{overpic}[trim=140 0 200 60, clip, width=0.49\textwidth]{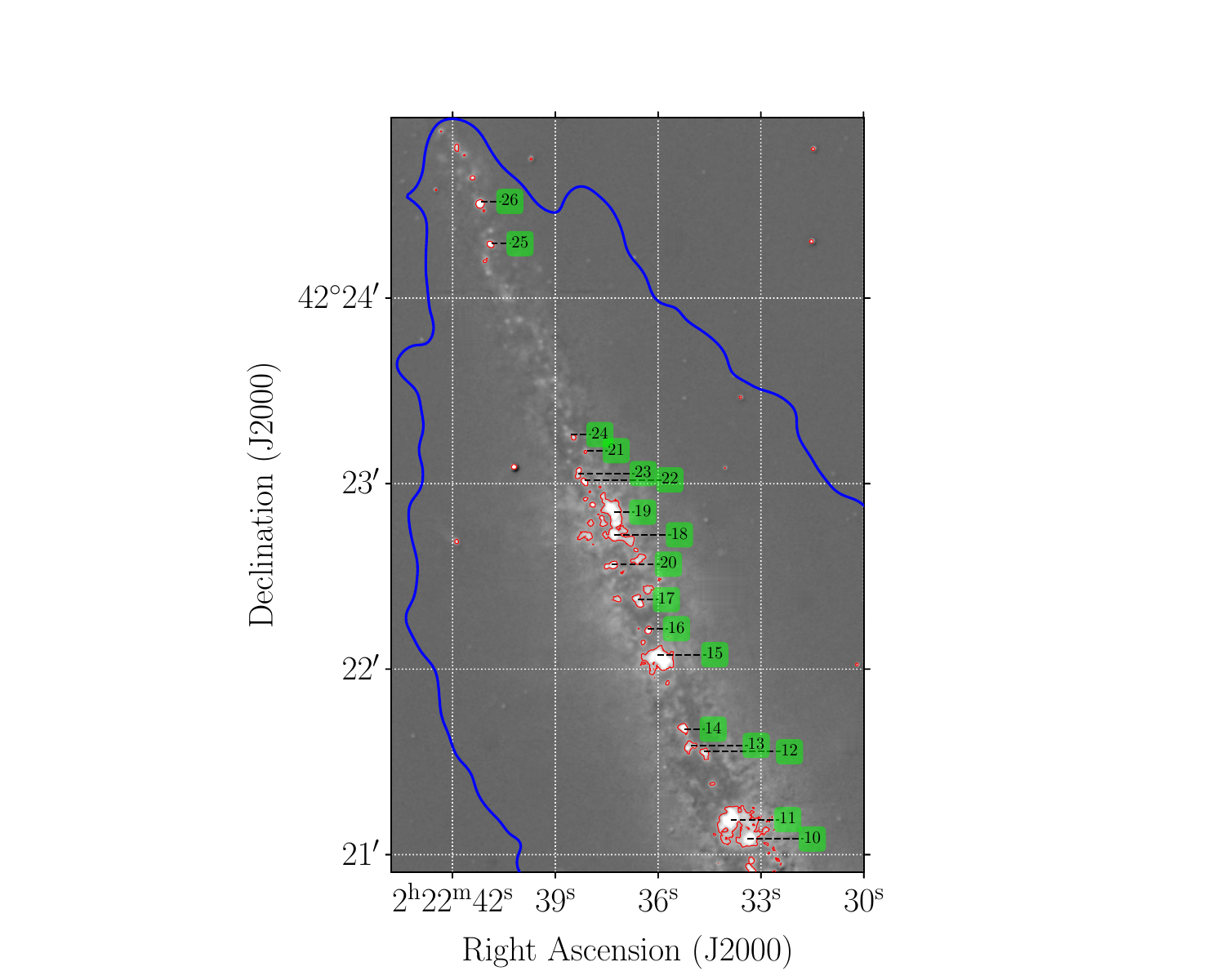}
        \put(65,93){\color{black}\setlength\fboxsep{1pt}\fcolorbox{black}{white}{\textbf{(b)}}}
    \end{overpic}
    \hfill
    \begin{overpic}[trim=140 0 200 60, clip, width=0.49\textwidth]{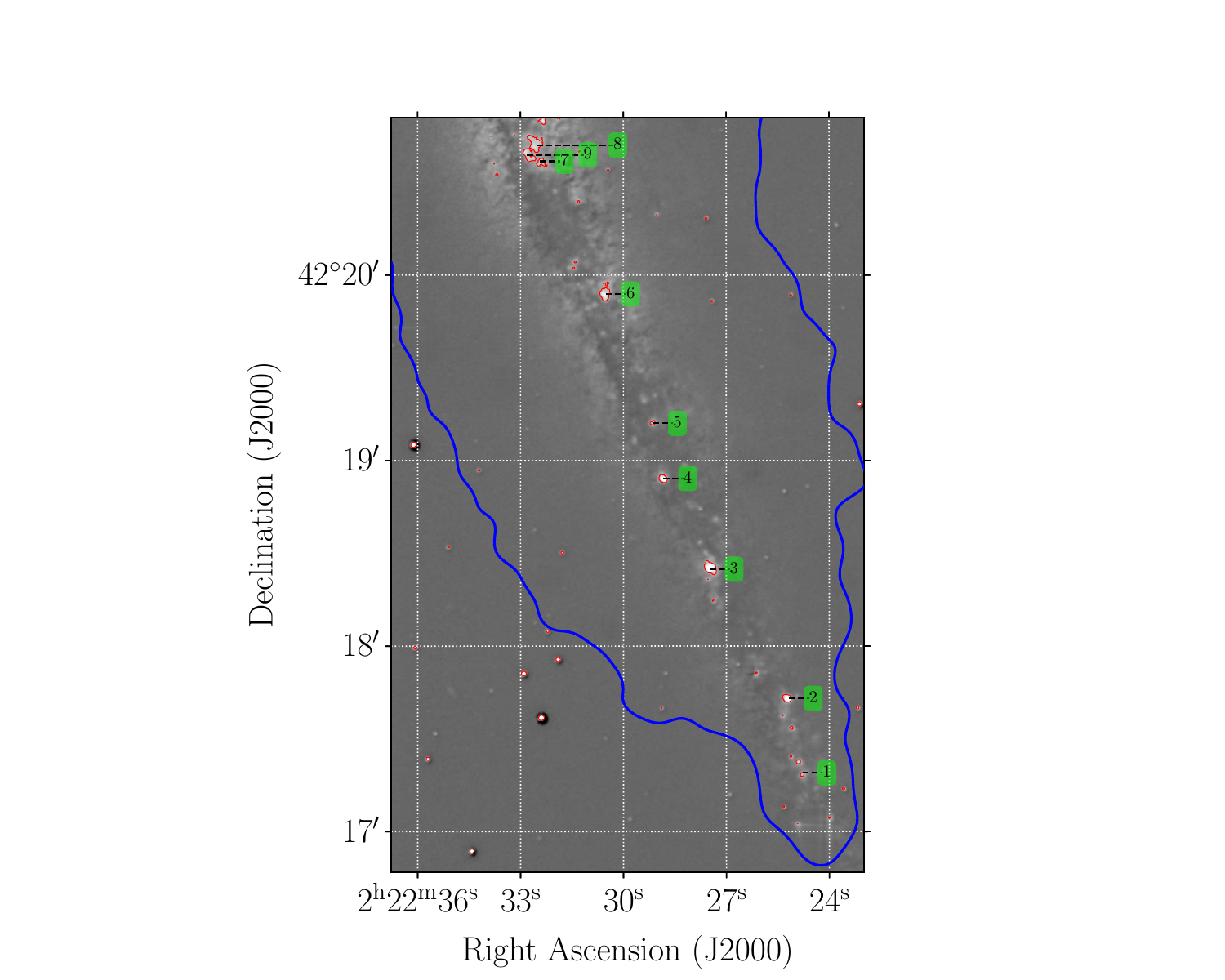}
        \put(65,93){\color{black}\setlength\fboxsep{1pt}\fcolorbox{black}{white}{\textbf{(c)}}}
    \end{overpic}

    \caption{a: Polar-coordinate map of the disk of NGC~891 showing the locations of H\,{\footnotesize II} regions identified from the $\rm H \alpha$ image. Points with negative position angles correspond to regions located in the north-east side of the galaxy.
    The red dashed curve indicates the radio continuum extent of the galaxy, defined here as $I \gtrsim 0.5\ \rm mJy\ Beam^{-1}$.
    b: $\rm H \alpha$ map of the north-east side of NGC~891 overlaid with a Stokes $I$ contour at $0.1\ \rm mJy\ Beam^{-1}$, highlighting H\,{\footnotesize II} regions from Table~\ref{tab:HIIregions}, indicated in green boxes. c: Same as b, but showing the south-west side of the galaxy.}
    \label{fig:H_alpha_maps}
\end{figure*}

\begin{figure}
    \centering

    \begin{overpic}[trim=100 0 230 80, clip, width=0.95\columnwidth]{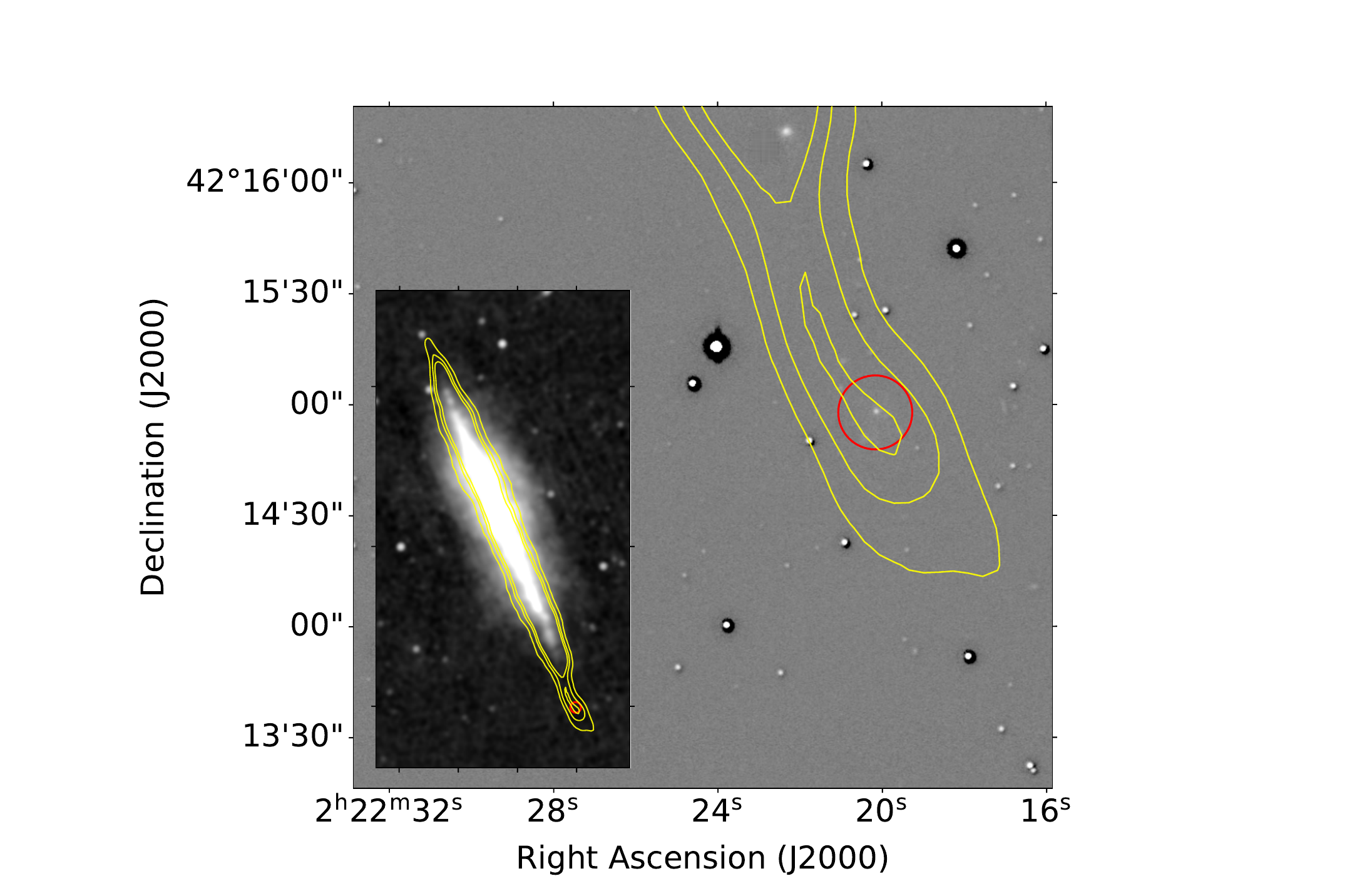}
    
        \put(90,78){\color{black}\setlength\fboxsep{1pt}\fcolorbox{black}{white}{\textbf{(a)}}}
    \end{overpic}
    \vspace{2mm}

    \begin{overpic}[trim=30 0 160 70, clip, width=0.95\columnwidth]{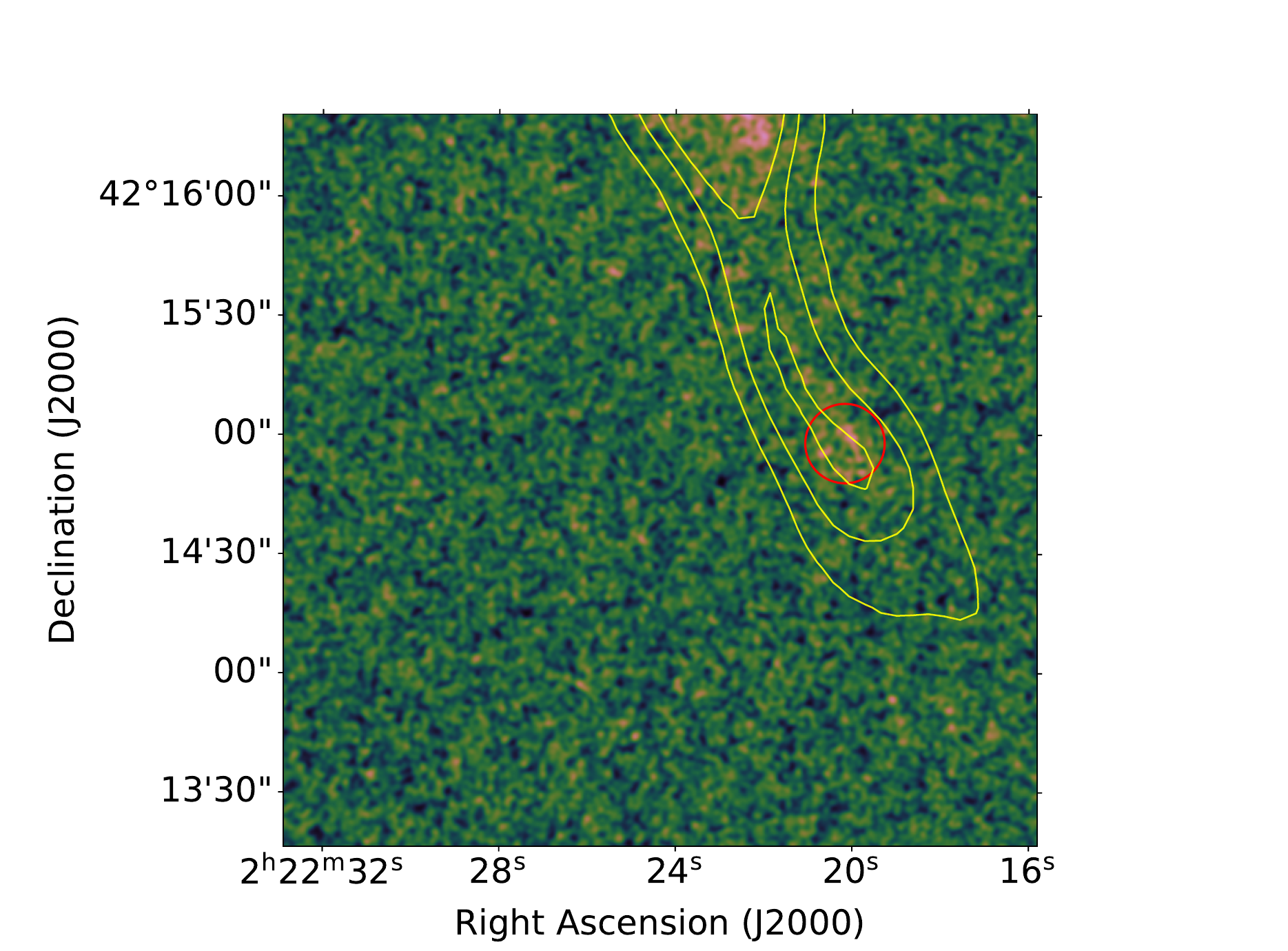}
        \put(90,77){\color{black}\setlength\fboxsep{1pt}\fcolorbox{black}{white}{\textbf{(b)}}}
    \end{overpic}

    \caption{Distant H\,{\footnotesize II} region in NGC~891 shown in a: $\rm H\alpha$ and b: far-UV, with yellow H\,{\footnotesize I} column density contours ($0.89,\ 1.5,\ 2.1\ \times10^{21}\ \rm cm^{-2} $) overlaid. In panel a, a lower left inset shows the Stokes $I$ radio continuum map (FWHM = 10\arcsec) of NGC~891 with the same H\,{\footnotesize I} contours, indicating the location of the distant H\,{\footnotesize II} region within the galaxy.}
    \label{fig:farHII}
\end{figure}

The outcomes of the three-dimensional RM synthesis and RM clean procedures are presented in Figure~\ref{fig:ngc891_3d_rm_synthesis_grid}, which displays a grid of twelve FDF maps at selected Faraday depths. Each panel corresponds to a slice through the FDF cube, separated by intervals of $100\ \rm rad\ m^{-2}$, spanning from $-650\ \rm rad\ m^{-2}$ to $+450\ \rm rad\ m^{-2}$. Overlaid white contours indicate the Stokes $I$ emission at the $18\sigma_I$ level, outlining the disk of NGC~891 and aiding in the identification of regions with significant total intensity emission. Cyan contours highlight areas within the FDF cube that exceed the $8\sigma$ detection threshold, where $\sigma = 3.3 \times10^{-6}\ \rm Jy\ Beam^{-1}\ RMTF^{-1}$ is the noise level in the real or imaginary component of the FDF, identifying locations of significant polarized emission. Although the sensitivity to large scale structure varies with wavelength, the structure in Stokes Q and U images should be broken up into smaller scales because of structure in polarization angle. We therefore do not expect to miss major structures in polarization. 

The $8\sigma$ detection threshold ensures that it is very unlikely that noise peaks in individual Faraday depth spectra lead to false detections. 
We also searched the Faraday depth spectra for peaks above this threshold within the range $-1500 \leq \phi \leq +1500\ \rm rad\ m^{-2}$.
Figure~\ref{fig:ngc891_3d_rm_synthesis_grid} shows extended structure below the $8\sigma$ threshold that is most likely real, polarized emission with a low surface brightness. It is possible to lower the detection threshold in RM synthesis when considering adjacent independent Faraday depth spectra, but that analysis is beyond the scope of this paper. We will include, where appropriate, a $6\sigma$ contour in some figures.

The first robust detection of polarized emission occurs at the Faraday depth of $-630\ \rm rad\ m^{-2}$, while the last appears at $+365\ \rm rad\ m^{-2}$. Although the displayed panels cover the full $-650$ to $+450\ \rm rad\ m^{-2}$ range, most of the strong, spatially extended polarized emission is concentrated between approximately $-150\ \rm rad\ m^{-2}$ and $+150\ \rm rad\ m^{-2}$. 
A polarized background radio source is detected through the halo of NGC~891 at RA = $02^{\rm h}22^{\rm m}44\fs87$ and DEC = $+42\arcdeg24\arcmin54\farcs54$ with a Faraday depth of $-176.2\ \pm 2.5\ \rm rad\ m^{-2}$.
Notably, the faint emission detected at the extreme negative end ($-650\ \rm rad\ m^{-2}$) is a counterpart of the emission feature at the high positive end ($+450\ \rm rad\ m^{-2}$).

To further investigate the nature of these polarized features detected at the extreme ends of the Faraday depth range, we present Figure~\ref{fig:ngc891_fdf_spectrum_3peaks_UVIT}. 
Figure~\ref{fig:ngc891_fdf_spectrum_3peaks_UVIT}a 
illustrates the clean FDF spectrum at $RA = 02^{\rm h}22^{\rm m}34\fs95$ and $DEC = +42\arcdeg 21\arcmin 41\farcs20 $ with three peaks exceeding the detection threshold, located at Faraday depths of $-630$, $-45$, and $+365\ \rm rad\ m^{-2}$. This region coincides with bright diffuse extraplanar $\rm H\alpha$ emission in \citet{rossa2004hubble}, including their filaments Fil-WF3-3/4 and Fil-WF3-5. The RMTF, shown in green and centered at the main peak ($-45\ \rm rad\ m^{-2}$), demonstrates that the neighboring peaks are neither symmetric nor mirror images of the sidelobe structure expected from the RMTF. This suggests that the features at $-630$ and $+365\ \rm rad\ m^{-2}$ are not residual sidelobes, but rather represent distinct Faraday-rotating components along the line of sight.

Additionally, far-UV images of NGC~891 are shown in grayscale for the same region, overlaid with polarized intensity contours at two extreme Faraday depths: Figure~\ref{fig:ngc891_fdf_spectrum_3peaks_UVIT}b corresponds to $\phi = -630\ \rm rad\ m^{-2}$ and Figure~\ref{fig:ngc891_fdf_spectrum_3peaks_UVIT}c corresponds to $\phi = 365\ \rm rad\ m^{-2}$. Yellow contours indicate the Stokes $I$ emission, tracing the galaxy’s disk, while blue contours represent polarized intensity at $6\sigma$, $8\sigma$, and $10\sigma$. 
The $6\sigma$ contour is shown in the darkest blue, the $8\sigma$ contour is drawn with the thickest line in an intermediate blue, and the $10\sigma$ contour is shown in the lightest blue. Red contours show $\rm H\alpha$ emission, highlighting H\,{\footnotesize II} regions listed in Table~\ref{tab:HIIregions}. The complex Faraday depth spectrum overlaps with H\,{\footnotesize II} regions 12-14. It appears bounded by H$\alpha$ filaments Fil-WF3-5 in the north and Fil-WF3-3/4 \citep{rossa2004hubble} in the south. The identification of the stars in the far-UV image indicates that the interstellar extinction to these H\,{\footnotesize II} regions is quite low. These stars appear in projection on the well-known dust lane of NGC~891.

Another notable region exhibiting complex Faraday rotation exists near the superbubble WF3-1-SS1 described by \citet{rossa2004hubble} ($RA = 02^{\rm h}22^{\rm m}33\fs602 ,\ DEC =+42\arcdeg 21\arcmin 10\farcs20$). This area is presented in detail in Figure~\ref{fig:ngc891_fdf_spectrum_2peaks_UVIT}.  Figure~\ref{fig:ngc891_fdf_spectrum_2peaks_UVIT}a displays the clean FDF spectrum at this position, revealing two peaks above the detection threshold at Faraday depths of $\phi = -55\ \rm rad\ m^{-2} $ and $\phi = 280\ \rm rad\ m^{-2}$, indicating the presence of multiple Faraday rotating components along the line of sight. Figure~\ref{fig:ngc891_fdf_spectrum_2peaks_UVIT}b and Figure~\ref{fig:ngc891_fdf_spectrum_2peaks_UVIT}c show the far-UV maps of this region in grayscale, overlaid with blue polarized intensity contours. At $\phi = 280\ \rm rad\ m^{-2}$ (Figure~\ref{fig:ngc891_fdf_spectrum_2peaks_UVIT}c), the region itself exhibits polarized emission, while the surrounding areas appear devoid of significant signal. In contrast, at $\phi = 120\ \rm rad\ m^{-2}$ (Figure~\ref{fig:ngc891_fdf_spectrum_2peaks_UVIT}b), the opposite is observed: polarized emission is detected in the surrounding area, but not within the target location.

Figure~\ref{fig:ngc891_polarization_maps} shows percentage polarization (Figure~\ref{fig:ngc891_polarization_maps}a), polarized intensity (PI) map with polarization E vector orientations corrected for Faraday rotation (Figure~\ref{fig:ngc891_polarization_maps}b), Rotation Measure (RM) overlaid with E vector orientations corrected for Faraday rotation(Figure~\ref{fig:ngc891_polarization_maps}c), and RM uncertainty (Figure~\ref{fig:ngc891_polarization_maps}d), all derived from RM-synthesis using the combined S-band and C-band data spanning $2.52–6.96\ \rm GHz$. The percentage polarization map shown in Figure~\ref{fig:ngc891_polarization_maps}a was derived using the synchrotron-only Stokes I cube, after subtraction of the thermal emission contribution. The polarized intensity map was constructed by selecting, for each pixel in $RA-DEC$ space, the maximum PI value along the Faraday depth cube that exceeded our detection threshold of $8\sigma$. As such, the reference frequency for polarized intensity is $3.26\ \rm GHz$, corresponding to the mean of $\lambda^2$ in the data. No attempt to correct for polarization bias was made because the signal to noise ratio in the image varies and the detection threshold at $8\sigma$ ensures the polarization bias is small.

The rotation measure (RM) map is a peak-RM map, in which each pixel is assigned the RM value corresponding to the location of the maximum PI in the Faraday depth cube. The RM has been corrected for foreground Galactic Faraday rotation of $-55.7 \pm 20\ \rm rad\ m^{-2}$ (Stil et al. in prep), which is less than the $-71\pm 10\ \rm rad\ m^{-2}$ derived by \citet{hutschenreuter2022galactic} at the $2\sigma$ level. The reason of the revision is an analysis that includes background sources in the NGC~891 field that indicate a lower Galactic RM. The revised foreground also uses new RM data from the SPICE-RACS survey \citep[Spectra and Polarization in Cutouts of Extragalactic sources from Rapid ASKAP Continuum Survey,][]{Thomson2023, Thomson2026}. The difference with \citet{hutschenreuter2022galactic} is consistent with the amplitude of Galactic RM structure on angular scales $\lesssim 1\degr$ that is undersampled in the RM catalog used by \citet{hutschenreuter2022galactic}. 

A comparison with the C-band polarization maps presented in Figure~5 of \citet{krause2020chang} shows that the polarized intensity distribution and Faraday-rotation corrected polarization E vector orientations are consistent, although our S-band polarized intensity map is more extended. The RM distribution, however, appears different. This difference is explained by the adopted correction for the Galactic foreground Faraday rotation. While \citet{krause2020chang} applied a foreground correction of $-80 \pm 18\ \rm rad\ m^{-2}$. Using the foreground correction adopted by \citet{krause2020chang} produces an RM map similar to their result, but the S-band RM map extends further into the halo.

The polarized emission that exceeds the detection threshold occurs primarily within the $18\sigma_I$ Stokes $I$ contour. The emission within the $100\sigma_I$ contour is notably less polarized than the emission farther from the galactic plane.
In Figure~\ref{fig:ngc891_polarization_maps}a, nine circles with the size of the synthesized beam mark the positions selected for analysis of depolarization as a function of wavelength.  
In Figure~\ref{fig:ngc891_polarization_maps}b and c, crosses indicate locations of notably high polarized intensity located along the galactic disk and large Faraday depth values; the corresponding clean FDF spectra for these regions are presented in Figure~\ref{fig:extreme_FDFs}.

Figure~\ref{fig:extreme_FDFs} shows the clean FDF spectra for two locations with notable features identified with crosses in Figure~\ref{fig:ngc891_polarization_maps} b and c. Figure~\ref{fig:extreme_FDFs}a corresponds to a region in the north-eastern halo of NGC~891 ($RA = 02^{\rm h} 22^{\rm m}34\fs684,\ DEC =+42\arcdeg 21\arcmin 50\farcs20$) characterized by high polarized intensity in the midplane, as seen in the PI map (Figure~\ref{fig:ngc891_polarization_maps}b). The FDF spectrum at this position reveals a single strong peak, indicating the presence of significant polarized emission at a single dominant Faraday depth. 

Figure~\ref{fig:extreme_FDFs}b displays the FDF spectrum for a region near the central part of the galaxy ($RA =02^{\rm h}22^{\rm m}32\fs700,\ DEC =+42\arcdeg 20\arcmin 59\farcs19$), selected for its unusually large Faraday depth ($\phi= 250\ \rm rad\ m^{-2}$), as indicated in the RM map (Figure~\ref{fig:ngc891_polarization_maps}c). The FDF profile in this central region shows a pronounced peak, suggesting the detected polarized emission arises from a single emitting region along the line of sight.  

\subsection{Magnetic field structure}

The magnetic field structure of NGC~891 is summarized in
Figure \ref{fig:AVG_RMprofileBvector}. Figure \ref{fig:AVG_RMprofileBvector}a shows the Faraday-rotation–corrected magnetic field orientation in the plane of the sky, overlaid with Stokes $I$ contours. The image is rotated by $-67.65\degr$ such that the projected major axis of the galaxy ($PA = 22.35\degr$) is horizontal and the northeast side of the galaxy is on the left. 
The pattern closely resembles the magnetic field structure derived by \citet{sukumar1991} at $6.2\ \rm cm$, without correction for Faraday rotation ($\sim 10\degr$ at 6 cm for $\rm |RM| = 60\ rad\ m^{-2}$). The magnetic field in the plane of the sky has a significant component perpendicular to the disk, leaving no compelling alignment with the major axis across most of NGC~891. The over-all structure resembles the X-shaped magnetic field structure reported by \citet{krause2008magnetic, krause2020chang}.

Figure~\ref{fig:AVG_RMprofileBvector}b presents the averaged RM profile, weighted by RM errors, along the projected major axis. In order to calculate the averaged RM profile, the RM map (Figure \ref{fig:ngc891_polarization_maps}c) was subdivided into rectangular regions parallel to the minor axis, each extending one beam along the major axis. Only regions containing at least three synthesized beams with detected polarization were retained. The resulting RM profile shows a smooth variation of RM along the major axis, with nearly constant RM on either side of the center and a resolved transition region near the projected minor axis. 

The interpretation of this RM structure depends critically on the true (Galactic) foreground Faraday rotation. Only if a sign change in RM occurs near the apparent minor axis, can this RM structure be interpreted in terms of a large-scale axially symmetric magnetic field.
Adopting the Galactic foreground correction of $-80\pm 18\ \rm rad\ m^{-2}$ used by \citet{krause2020chang} or the revised $-71\pm 10\ \rm rad\ m^{-2}$ results in positive mean RM on both sides of the minor axis. The revised foreground is $-55.7 \pm 20\ \rm rad\ m^{-2}$, with the error including the uncertainty from unsampled small-scale structure in the Galactic foreground. The error is indicated by the gray area in Figure~\ref{fig:AVG_RMprofileBvector}b. This leaves the possibility of an RM sign change across the minor axis, although the Galactic foreground needs to be determined with higher precision for a definitive conclusion. 

Figure~\ref{fig:AVG_RMprofileBvector} displays an interesting conundrum. The plane-of-sky magnetic field orientation corrected for Faraday rotation does not show a magnetic field predominantly oriented along the galactic plane, but the line-of-sight component, traced by Faraday rotation, shows a pattern consistent with a large-scale axially symmetric magnetic field. The two observations are not mutually exclusive because the plane of sky component of the magnetic field still has a component parallel to the galactic plane. Its direction is toward the north east, if the large-scale RM pattern is made by an axially symmetric magnetic field.

\subsection{Wavelength dependence of depolarization}

The pattern observed in the percentage polarization map (Figure~\ref{fig:ngc891_polarization_maps}a) is consistent with what has been reported in other edge-on galaxies \citep{krause2020chang}: the thin disk exhibits a lower degree of polarization, whereas the halo shows higher fractional polarization. This pattern may arise from wavelength-dependent depolarization related to differential Faraday rotation across the beam and along the line of sight. It may also include wavelength-independent depolarization from unresolved magnetic field structures that change with distance along the line of sight.

To further investigate the wavelength dependence of fractional polarization, Figure~\ref{fig:fracpol2_lambda2} presents graphs of the \textit{squared} fractional polarization as a function of $\lambda^2$ for nine representative locations marked by colored circles in Figure~\ref{fig:ngc891_polarization_maps}a. The fractional polarization was calculated using the synchrotron-only Stokes $I$ cube described in Section \ref{sec: RM synthesis}, after subtraction of the thermal emission contribution. We consider the square of fractional polarization because the polarization bias correction is linear in this quantity. No attempt was made to correct for polarization bias as the $8\, \sigma$ detection threshold ensures that the effect of polarization bias is small.

The dots in Figure~\ref{fig:fracpol2_lambda2} represent the data at the nine positions marked in Figure~\ref{fig:ngc891_polarization_maps}a. The crosses indicated in Figure~\ref{fig:fracpol2_lambda2} represent the expected median values of $p_\lambda^2$ if there is no signal. This median is calculated from the Rayleigh distribution for $p_\lambda$ in the absence of signal, transformed to the probability density function for $p^2_\lambda$ and integrated. The median value from noise only is $(2 \ln 2)\, n_\lambda^2 \approx 1.39\,n_\lambda^2$, where $n_\lambda^2=\sigma_Q^2/I_\lambda^2$, $\sigma_Q = \sigma_U$ is the standard deviation of the noise in Stokes $Q$ or Stokes $U$, and $I_\lambda$ is the total intensity at wavelength $\lambda$ for the position of interest. The noise spectra show much less scatter than the data because they represent the standard deviation over a large solid angle with no detectable polarized emission, while the data represent a single beam.  

The $p_\lambda^2$ values in the halo and in the disk exceed the median of the noise, except in the central panel of Figure~\ref{fig:fracpol2_lambda2}, which coincides with a position with no signal above the detection threshold (Figure~\ref{fig:ngc891_polarization_maps}a). 

As an additional check on the effect of noise, the $Q$ and $U$ images were averaged over four adjacent channels in the middle of S-band. 
Frequency averaging the data does not significantly lower $p_\lambda^2$, but it lowers the median of the noise by a factor $\sim 4$, which is the expected value if $\sigma_Q$ and $\sigma_U$ decrease by a factor 2 after averaging 4 independent channels. This experiment shows that polarization bias does not affect the slope of $p_\lambda^2$ as indicated by the data.

The horizontal dashed lines correspond to percentage polarization of $10\%$ and $1\%$. Figure~\ref{fig:fracpol2_lambda2} shows that wavelength-dependent depolarization is at most a factor $\sim 10$ in $p_\lambda^2$, which corresponds to a depolarization factor $\sim 1/3$ in fractional polarization over the $\lambda^2$ range of C-band and S-band combined.   
We fitted the data in Figure~\ref{fig:fracpol2_lambda2} with the internal Faraday dispersion model of \citet{sokoloff1998depolarization} (their Eq. 34), which describes the complex polarization as
\begin{equation}
    \mathcal{P} = p_0 \frac{1 - \exp(-S)}{S},
\end{equation}
where $S= 2 \sigma_\phi^2 \lambda^4 - 2i \mathcal{R}\lambda^2$, and $\mathcal{R}$ is the total Faraday depth of the layer, $\sigma_\phi$ is the Faraday depth dispersion, and $p_0$ is the intrinsic polarization at $\lambda = 0$. For the rotation measure of this model we have $|\rm{RM}| \leq \frac{1}{2} |\mathcal{R}|$, where the equality holds when $\sigma_\phi = 0$ (differential Faraday rotation only). For a combination of differential Faraday rotation and Faraday dispersion ($|\mathcal{R}| > 0$ and $\sigma_\phi>0$), the far side of the layer is selectively more depolarized. Note that in this case the RM of the model depends on wavelength. Experiments with RM synthesis on this model indicated that $\mathcal{R} = 70\ \rm rad\, m^{-2}$ yielded RM $\approx 22\ \rm rad\, m^{-2}$ for $\sigma_\phi$ in the range that fits the data. The resulting $\sigma_\phi$ are only mildly sensitive to the adopted value of $\mathcal{R}$. The squared modulus of the model, $|\mathcal{P}(\lambda^2)|^2$, was fitted to the observed $p_\lambda^2$ data, and the best-fit values of $\sigma_\phi$ and $p_0$ are indicated in each panel of Figure~\ref{fig:fracpol2_lambda2}.

The depolarization in the galactic plane shows modest wavelength dependence with weak wavelength dependence in the halo. This is remarkable because the density of the warm ionized medium is expected to decrease significantly over the vertical distance probed in Figure~\ref{fig:fracpol2_lambda2}.

\begin{figure*}
    \centering

    \begin{minipage}[t]{0.48\textwidth}
        \centering
        \begin{overpic}[trim=10 45 0 40, clip, width=\textwidth]{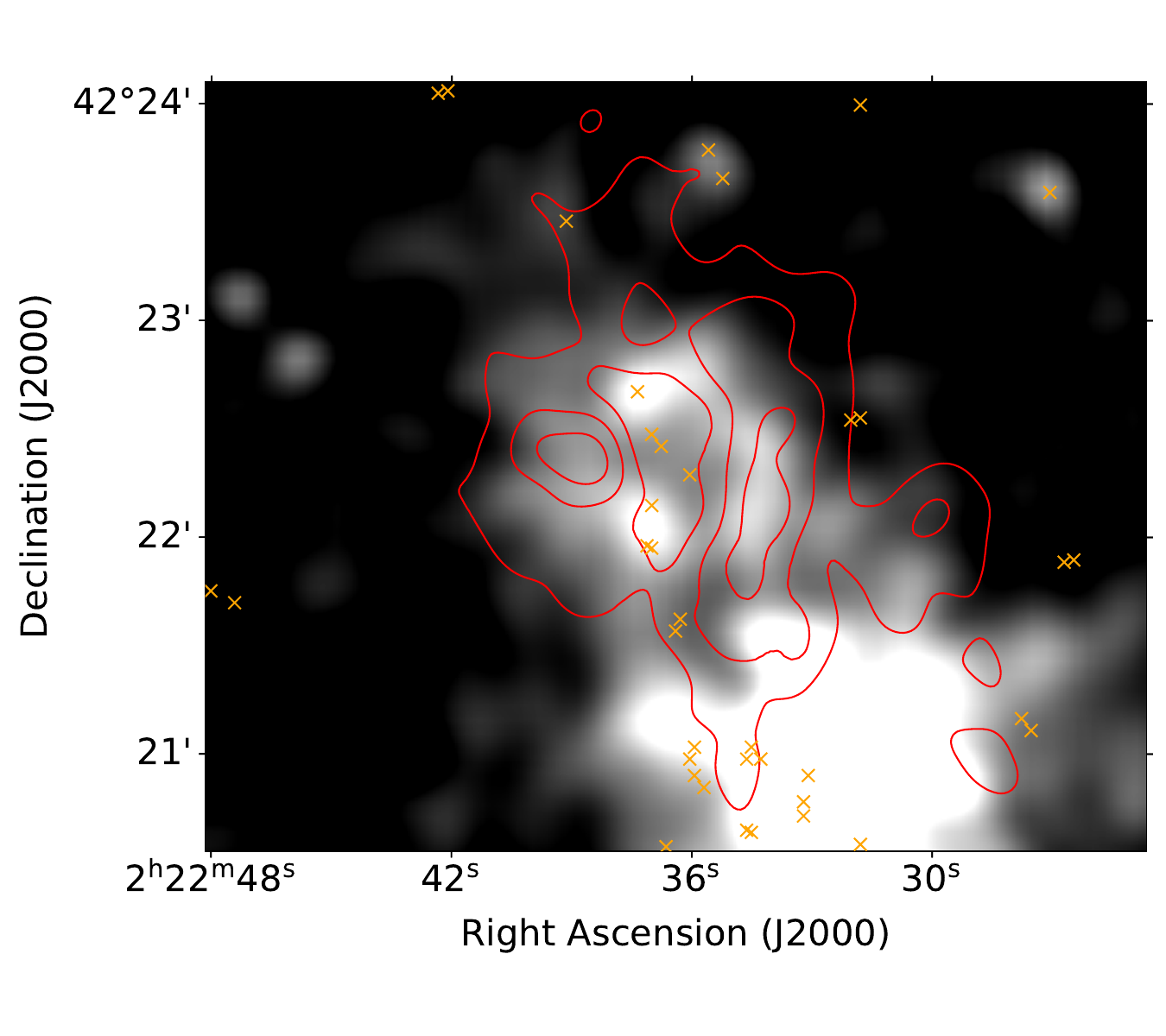}
            \put(20,70){\color{black}\setlength\fboxsep{1pt}\fcolorbox{white}{white}{\textbf{(a)}}}
        \end{overpic}
    \end{minipage}
    \hfill
    \begin{minipage}[t]{0.48\textwidth}
        \centering
        \begin{overpic}[trim=10 45 0 40, clip, width=\textwidth]{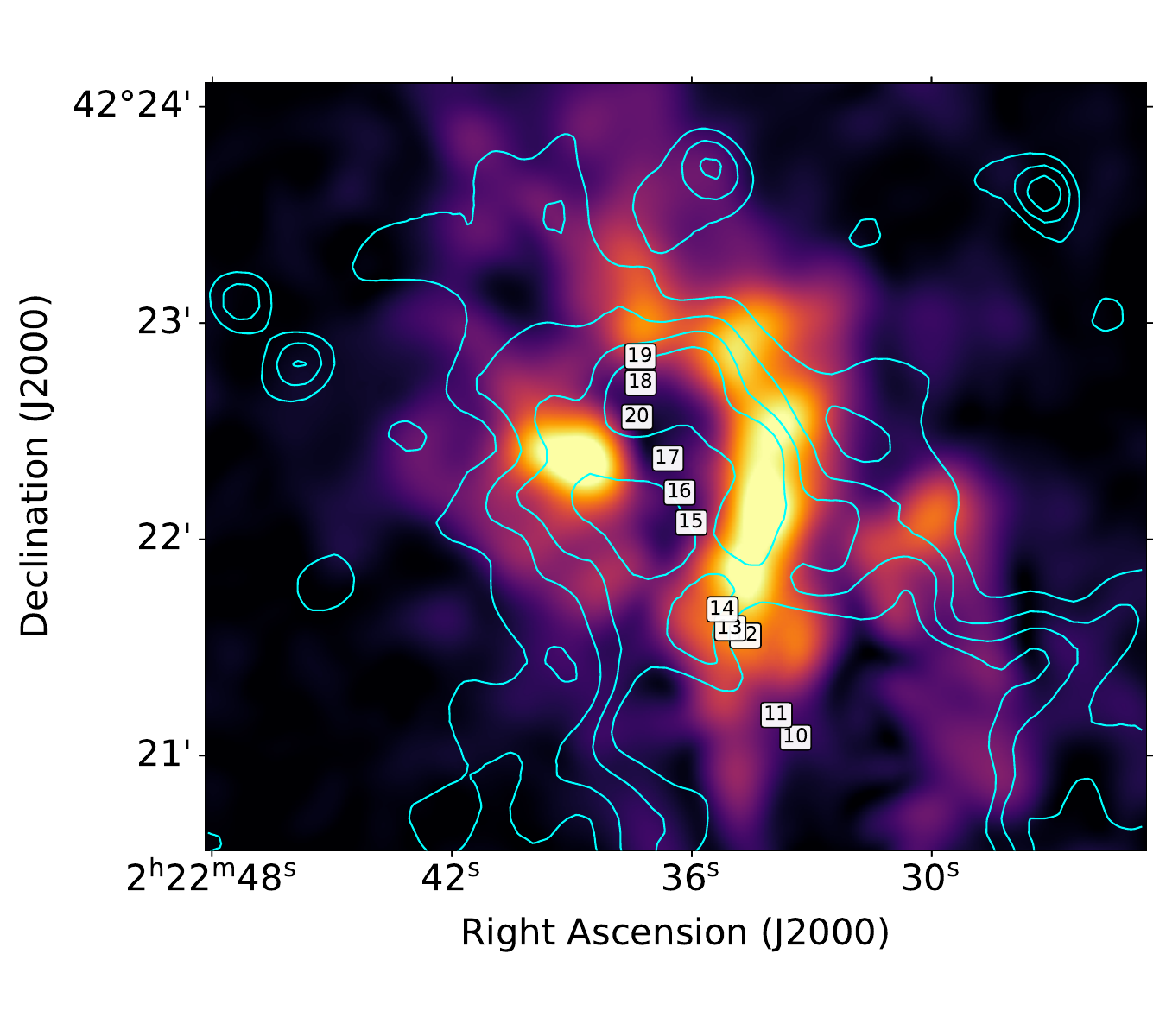}
            \put(20,70){\color{black}\setlength\fboxsep{1pt}\fcolorbox{white}{white}{\textbf{(b)}}}
        \end{overpic}
    \end{minipage}

    \caption{Comparison of X-ray and Faraday depth structures in NGC~891. 
a: X-ray emission map with contours of the cleaned FDF overlaid at levels of $5\times10^{-5}$, $1\times10^{-4}$, and $1.5\times10^{-4}\ \rm Jy\ Beam^{-1}$, highlighting polarized features coincident with X-ray structures. The crosses mark the locations of subtracted X-ray point sources.
b: FDF slice at $\phi=-75\ \rm rad\ \rm m^{-2}$ shown with X-ray contours at levels of 8, 9, 10, and 11 $\rm counts\ pix^{-1}$, emphasizing the spatial correspondence between regions of high polarized intensity and diffuse X-ray emission. The HII regions within this region are indicated by their labels.}

    \label{fig:Xray_FDF}
\end{figure*}

\subsection{Multi-wavelength results}
\subsubsection{\texorpdfstring{H$\alpha$ intensity}{H alpha intensity}}
We investigated the spatial distribution of H\,{\footnotesize II} regions in the disk of NGC~891 using the $\rm H\alpha$ spectral cube obtained from \citet{kamphuis2007}. From this dataset, we constructed a polar map of the galaxy by converting the positions of the detected H\,{\footnotesize II} regions into galactocentric coordinates, assuming circular orbits with a constant rotational velocity of $221.59\ \rm km\ s^{-1}$ \citep{fraternali2011tale} and using the following equation:
\begin{equation}
    v_{los}-v_{sys} = v_R\ \sin{\theta},
    \label{HIIregionsGeometry_eq}
\end{equation}
where $v_{los}$ is the line-of-sight velocity of an H\,{\footnotesize II} region (extracted from the $\rm H \alpha$ cube), $v_{sys}$ is the systemic velocity, $528\ \rm km\ s^{-1}$ \citep{oosterloo2007cold, fraternali2011tale}, $v_R$ is the rotational velocity, and $\theta$ is the angle between the line of sight and the galactocentric radius of an  H\,{\footnotesize II} region ($R$). For each H\,{\footnotesize II} region we use Equation \ref{HIIregionsGeometry_eq} to solve for $\sin\theta$. The value of $R$ is calculated as the projected distance of an H\,{\footnotesize II} region from the galaxy center, divided by $\sin\theta$. Strictly speaking there is an ambiguity between points in the near side and the far side of NGC~891. However, interstellar extinction will render H\,{\footnotesize II} regions on the far side undetectable, which resolves the near-far ambiguity.

Figure~\ref{fig:H_alpha_maps} presents the results of this analysis. Figure~\ref{fig:H_alpha_maps}a shows the face-on map of the H\,{\footnotesize II} regions, while Figure~\ref{fig:H_alpha_maps}b and Figure~\ref{fig:H_alpha_maps}c display the $\rm H\alpha$ emission maps \citep{vargas2018chang} of the north-east and south-west sides of the galactic disk, respectively. H\,{\footnotesize II} regions are labeled according their number in Table~\ref{tab:HIIregions}, and the galaxy's disk is outlined with blue Stokes $I$ contours at the $0.1\ \rm mJy\ Beam^{-1}$ level to highlight the extent of the radio-emitting region. Red contours mark $\rm H\alpha$ intensity at $1.5 \times 10^{-17}\ \rm erg\ s^{-1}\ cm^{-2}\ pix^{-1}$. 
H\,{\footnotesize II} regions 7, 8, and 9 were excluded from the face-on map and subsequent calculations, as their line-of-sight velocities are within $20\ \rm km\ s^{-1}$ of the $v_{sys}$. The radius $R$ is not constrained for H\,{\footnotesize II} regions toward the center of NGC~891 if $v_{los} \approx v_{sys}$.

The face-on map reveals a noticeable asymmetry in the radial distribution of H\,{\footnotesize II} regions between the two sides of the galaxy. On the north-east side, H\,{\footnotesize II} regions 15-24 are located closer to the galactic center, whereas those on the south-west side are distributed farther out. 
This asymmetry is consistent with the trailing spiral structure of NGC~891 in \citet{kamphuis2007dust}, although a fit of a logarithmic spiral to the H\,{\footnotesize II} regions 15 to 26 in the northeast and 1 to 5 in the southwest suggested these H\,{\footnotesize II} regions are not all on the same spiral arm.
The red dashed circle in  Figure~\ref{fig:H_alpha_maps}a indicates the radio continuum size of the galaxy, which corresponds to the extent of the region where $I \gtrsim  0.5\ \rm mJy\ Beam^{-1}$. Also, near-infrared (3.6 $\mu$m) surface photometry shows a rapid decrease in surface brightness beyond $250\arcsec$ (11 kpc) from the center of NGC~891 \citep{fraternali2011tale}, even though the disk extends to at least $400\arcsec$ (18 kpc) from the center. 

Two sources of uncertainty affect these calculations: the adopted rotational velocity and the line-of-sight velocity of each H\,{\footnotesize II} region. 
The uncertainty in rotational velocity was assessed using the values tabulated in \citet{fraternali2011tale} by first computing their weighted mean and then identifying the value with the largest deviation from this mean. Both the weighted mean and this extreme value were used separately in the calculations, and the resulting difference was taken as the rotational velocity uncertainty. For the line-of-sight velocities, uncertainties were evaluated by adding/subtracting half of the velocity-channel width of the $\rm H\alpha$ cube to the measured values, and repeating the calculations with these perturbed velocities. The total radial error is a combination of these two contributions. 
The resulting uncertainties in $R\cos{\theta}$ range from approximately $0.45\ \rm kpc$ to $2.29\ \rm kpc$, while the uncertainty in $R \sin \theta$ is negligible. The uncertainties in $R \cos \theta$ are shown as the vertical error bars in Figure~\ref{fig:H_alpha_maps}.

An intriguing feature revealed by our $\rm H\alpha$ analysis is the presence of an H\,{\footnotesize II} region located beyond the visible star-forming disk of NGC~891. This region is highlighted in Figure~\ref{fig:farHII}, which presents a multi-wavelength view of its environment. 
Figure~\ref{fig:farHII}a provides a zoomed-in view of this area in $\rm H\alpha$. 
An inset in the lower left of this panel shows the Stokes $I$ radio continuum map of NGC~891, indicating the location of the distant H\,{\footnotesize II} region within the galaxy. In the $\rm H\alpha$ image, stars of similar brightness have a negative shadow because of imperfect continuum subtraction. H\,{\footnotesize II} regions do not have this shadow. Figure~\ref{fig:farHII}b displays the far-UV image of the same region, revealing spatially coincident UV emission. In both figures, the maps are overlaid with yellow H\,{\footnotesize I} column density contours at levels of $0.89,\ 1.5,\ 2.1\ \times10^{21}\ \rm cm^{-2} $, highlighting the distribution of atomic gas in the region. 
The presence of both $\rm H\alpha$ and UV emission at such a large galactocentric radius implies that the star-forming disk of NGC~891 may extend farther than previously assumed. Low-level star formation across the H\,{\footnotesize I} disk of galaxies is not unusual \citep[e.g.][]{ferguson1998discovery}. This H\,{\footnotesize II} region is interesting because its location supports the conclusion from Figure~\ref{fig:H_alpha_maps} that the visible H\,{\footnotesize II} regions in the southwest side of NGC~891 are located at large galactocentric radius. 

\section{Discussion} \label{sec:Discussion}

\subsection{North-South asymmetry}
Multi-resolution Stokes $I$ maps of NGC~891 (Figure~\ref{fig:Stokes_I_images}) reveal an extended radio halo enveloping the disk. \citet{heesen2025} fitted a thin disk with scale height 0.13 kpc and a thick disk with scale height of 1.00 kpc to these data. \citet{schmidt2019chang} found slightly larger values for the L-band and C-band data. The radial extent of the continuum emission closely resembles the radial extent of the bright H\,{\footnotesize II} regions and the brighter stellar disk. The inner disk is notably brighter in the north-east. At larger heights the emission becomes more diffuse and exhibits protrusions in the far outer halo; these features are most evident in the $20\arcsec$ map. 
The morphology is consistent with the total intensity maps of NGC~891 in L-band and C-band presented by \citet{schmidt2019chang}.
A similar morphology is seen in the Spitzer MIPS $70\ \mu$m emission presented by \citet{whaley2009multiwavelength}, which traces cool dust distributed above the disk. In particular, the broad diffuse emission and the shell-like structure on the eastern side of NGC~891 resemble the extended features seen in our low-resolution Stokes $I$ map (Figure~\ref{fig:Stokes_I_images}c).

The most significant asymmetry in the halo is the sudden broadening of the Stokes $I$ contours in the north-east side coinciding with tracers of hot massive stars in the disk, as described previously by \citet{kamphuis2007dust}. The present work indicates that this is a localized feature in the outer disk of NGC~891 (Figure~\ref{fig:H_alpha_maps}).  
The notion of a region of localized enhanced star formation is also supported by the coincidence of significant diffuse X-ray emission that is otherwise only seen in the central region of NGC~891 \citep{hodges2018hot, temple2005x, strickland2004high, bregman1994x}. 
\citet{hodges2018hot} found that the diffuse X-ray emission averaged over the inner halo of NGC~891 arises from a cool (0.2 keV) thermal component, a hotter (0.7 keV) thermal component, and a nonthermal component closer to the center of NGC~891.    
Figure~\ref{fig:Xray_FDF} shows the relative locations of polarized emission and diffuse X-ray emission, co-adding archival Chandra observations (~\dataset[DOI:10.25574/cdc.601]{https://doi.org/10.25574/cdc.601})
 with point sources subtracted, and convolved to a $15\arcsec$ beam. This region is in the disk and inner halo as defined in \citet{hodges2018hot}.

\begin{figure*}
    \centering

    \begin{overpic}[trim=10 10 10 10, clip, width=0.95\textwidth]{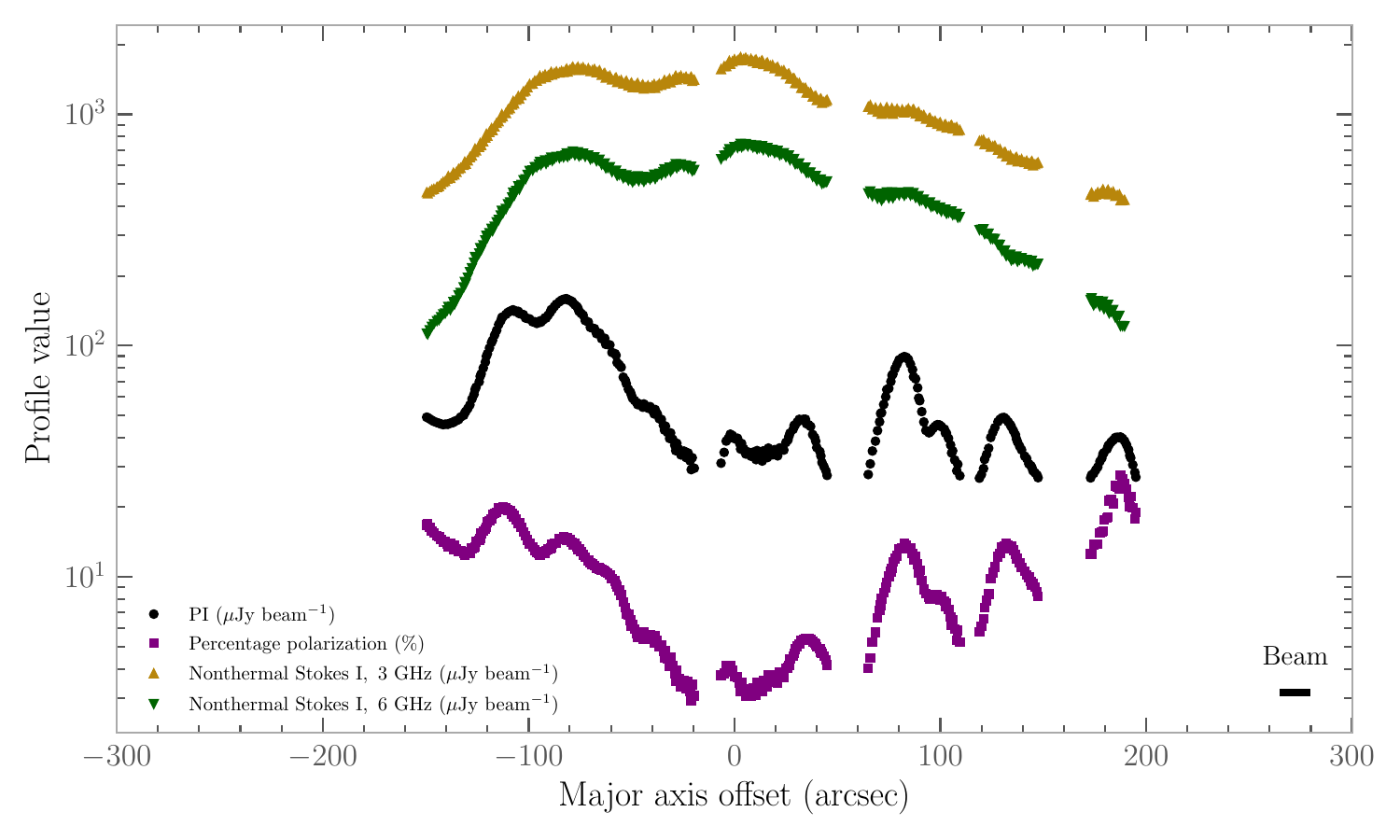}
        \put(10,55){\color{black}
        \setlength\fboxsep{1pt}
        \fcolorbox{black}{white}{\textbf{(a)}}}
    \end{overpic}

    \vspace{6mm}

    \begin{overpic}[trim=10 10 10 10, clip, width=0.95\textwidth]{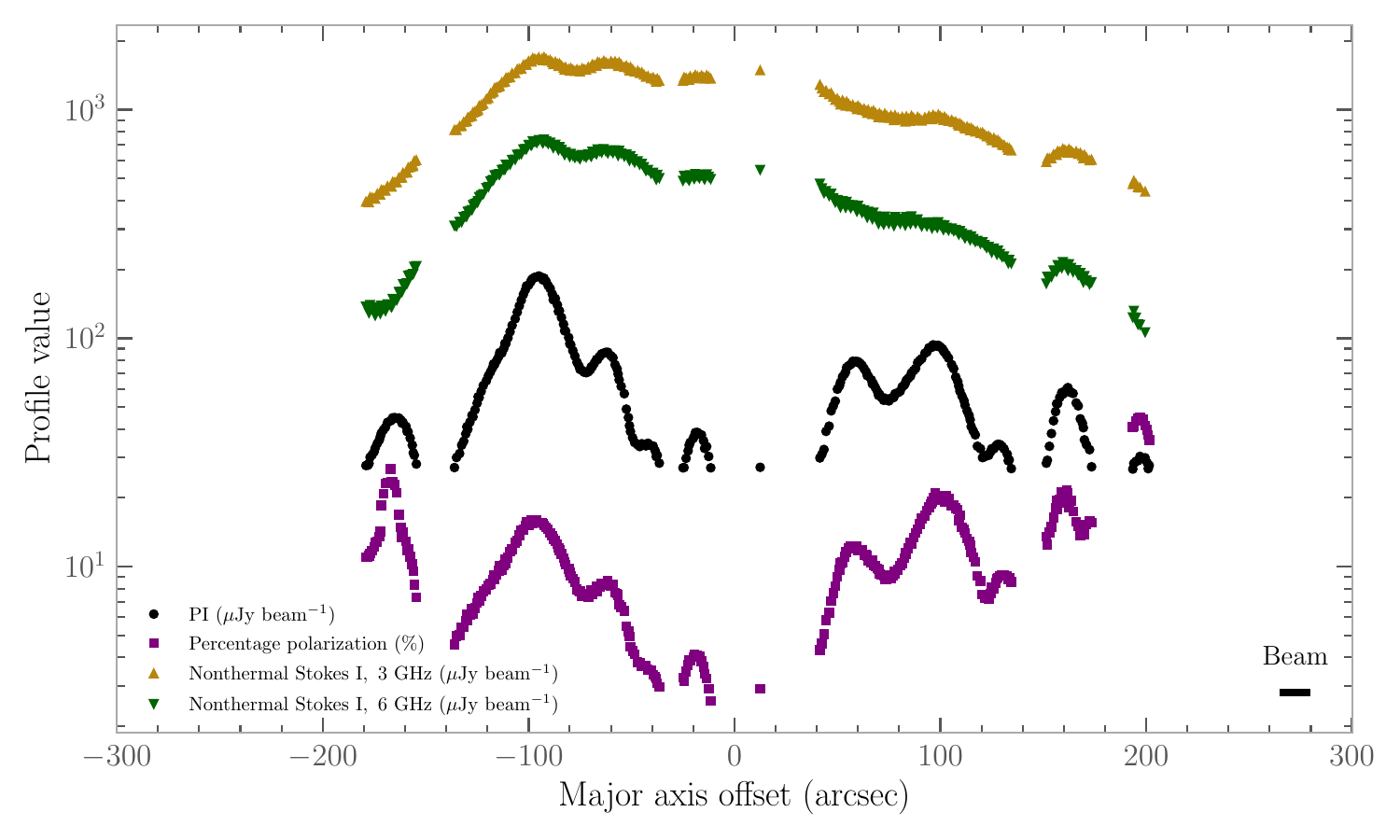}
        \put(10,55){\color{black}
        \setlength\fboxsep{1pt}
        \fcolorbox{black}{white}{\textbf{(b)}}}
    \end{overpic}

    \caption{ Profiles of polarized intensity (PI) at 3.26 GHz, percentage polarization at 3.26 GHz, nonthermal Stokes I at 3 GHz, and nonthermal Stokes I at 6 GHz as a function of major-axis offset in NGC~891. Panel a shows profiles measured along a line parallel to the major axis passing through RA = $02^{\rm h}22^{\rm m}33\fs87$ and DEC = $+42\arcdeg22\arcmin38\farcs20$, located on the right-hand side of the major axis and offset from it by two beam widths. Panel b shows profiles measured along a line parallel to the major axis passing through RA = $02^{\rm h}22^{\rm m}39\fs11$ and DEC = $+42\arcdeg22\arcmin20\farcs20$, located on the left-hand side of the major axis and offset from it by two beam widths. Negative major-axis offsets correspond to the northeastern side of the galaxy. The prominent high-polarization region discussed in the text is located at major-axis offsets of approximately $-110\arcsec$ to $-80\arcsec$. The beam size is indicated by the horizontal bar in the lower-right corner of each panel. All profiles are shown on a logarithmic scale. The percentage polarization profile is calculated using the nonthermal Stokes I emission.
    }

    \label{fig:IntensityProfiles}
\end{figure*}

Figure \ref{fig:IntensityProfiles} shows nonthermal total intensity, polarized intensity and percentage polarization as a function of position along lines parallel to the major axis at offsets $30\arcsec$ ($1.32\ \rm kpc$) on either side of the midplane. The highest polarized intensity occurs at major axis offsets of $-81.5\arcsec$ in Figure \ref{fig:IntensityProfiles}a and $-95\arcsec$ in Figure \ref{fig:IntensityProfiles}b. At these locations we also see a higher nonthermal total intensity at 3 GHz and at 6 GHz, compared to the equivalent positions in the SW offsets $81.5\arcsec$ and $95\arcsec$, respectively . This corresponds to the widening of the Stokes I contours in Figure \ref{fig:ngc891_polarization_maps}b. The excess total intensity at 3 GHz relative to the corresponding point in the southwest is $523\ \rm \mu Jy\ Beam^{-1}$ for Figure \ref{fig:IntensityProfiles}a and $768\ \rm \mu Jy\ Beam^{-1}$ for Figure \ref{fig:IntensityProfiles}b. A similar subtraction is more complicated in polarization because of its vector nature. We calculated the mean of the difference between the northeast side and the southwest side over all possible angles. This results in percentage polarization of $33\%$ and $26\%$ for Figure \ref{fig:IntensityProfiles} a and b, respectively.   

The association of this synchrotron emitting structure with the diffuse X-rays (Figure \ref{fig:Xray_FDF}) and therefore H\,{\footnotesize II} regions 15 to 20 locates this structure on the Earth-facing side of NGC~891, near the edge of the optical/radio disk.
\citet{sukumar1991} identified this region in 6 cm observations of NGC~891 as an anomalous feature approximately 3 kpc northeast of the nucleus, characterized by highly aligned magnetic fields crossing the galactic disk. Figure \ref{fig:AVG_RMprofileBvector}a shows a similar ordered magnetic field pattern at the same location.

\citet{rossa2004hubble} identified a supergiant shell WF3-3-SS2 and a supershell WF4-SS1 with H\,{\footnotesize II} regions 15-20 (see Table~\ref{tab:HIIregions}). 
These supershells have dimensions of $700 \times 511$ pc (WF3-3-SS2) and $345 \times 244$ pc (WF4-SS-1). Both these supershells are contained within the $\sim 2\ \rm kpc$ region with enhanced polarized intensity (Figure \ref{fig:Xray_FDF}).
The interior of such a supershell contains low-density hot, ionized gas visible in soft X-rays, conditions that would minimize the effects of Faraday rotation and thereby reduce depolarization, allowing polarized emission to remain clearly detectable. In contrast, the surrounding regions are likely to consist of cooler and denser plasma, where enhanced Faraday depolarization suppresses the polarized signal, producing either very faint or no polarized emission \citep[e.g.][]{west2007,gao2015}. 
 
Radio continuum of Galactic superbubbles tends to be thermal bremsstrahlung and superbubbles are modeled as passive Faraday screens with no internal synchrotron emission \citep{gao2015,thomson2018}. A mechanism for significant synchrotron emission from an evolved superbubble is required if association of the strong polarized emission in NGC~891 with a superbubble or a complex of supershells is true. Particle acceleration at the outer shock of a superbubble appears unlikely because of the small shock velocity, but there is a strong case for cosmic ray acceleration in compact star clusters inside the superbubble, e.g. \citet{rocamora2025} and references therein, and as a source of GeV gamma rays \citep{abdo2010}. The cosmic ray production and retention in superbubbles was modeled by \citet{vieu2022}. The balance of cosmic ray acceleration and escape from the superbubble may determine the visibility of synchrotron emission. Containment of cosmic ray electrons within the superbubble may depend on local circumstances.
In the case of NGC~891, we identify H\,{\footnotesize II} regions 15 to 20 with the superbubble, but we have no further information about star clusters inside the superbubble. 

The superbubble 30 Dor C in the Large Magellanic Cloud, shows shell-like radio emission and X-ray emission, both of which include non-thermal spectral components \citep{smith2004confronting, kavanagh2015, mayer2025}. This superbubble is powered by the OB association LH 90 that contains clusters with ages 3 Myr to 7 Myr \citep{testor1993}. 

\citet{west2021unified} argued that most of the extended diffuse Galactic polarization observed from Earth is part of a single local superbubble/chimney filled with soft-X-ray-emitting gas and surrounded by cold neutral gas and dust \citep{pelgrims2020modeling}.  
But for the Galactic Fan Region, there is no evidence for excess Stokes $I$ emission compared with the surrounding parts of the Galaxy \citep{hill2017fan}, whereas in NGC 891 we detect an enhancement in Stokes $I$. Moreover, the spatial extent of the feature in NGC 891 ($\sim 2\ \rm kpc$) appears unusually large for a single superbubble.

\subsection{Depolarization}
\label{sec:depol_discuss}
The low fractional polarization of the disk of NGC~891 compared to the halo is consistent with C-band observations of other galaxies in the CHANG-ES survey \citep{krause2020chang}. The results presented in Figure~\ref{fig:fracpol2_lambda2} show very mild wavelength-dependent depolarization in this particular region of NGC~891. This weak wavelength dependence in the disk and in the halo is an important clue to the nature of the depolarization. The dimensionless quantity $\sigma_\phi \lambda^2$ is a measure of wavelength-dependent depolarization by differential Faraday rotation and internal Faraday dispersion, together quantified by the Faraday depth dispersion $\sigma_\phi$. When $\sigma_\phi \lambda^2 \gtrsim 1$, significant depolarization is expected because relative polarization angle changes $\gtrsim 1\ \rm rad$ occur within the volume probed by the beam. In this regime, depolarization depends strongly on wavelength, even if the details are model dependent \citep[e.g.][]{burn1966depolarization,tribble1991depolarization,sokoloff1998depolarization,shanahan2023,stil2025}. 

To put the results of the fits in Figure~\ref{fig:fracpol2_lambda2} into perspective, what value for $\sigma_\phi$ should be expected for the complete edge-on disk of NGC~891? The $15\arcsec$ beam of our polarization data corresponds to a scale of 0.66 kpc. The expected Faraday depth dispersion within the beam can be estimated from the RM dispersion of extragalactic sources in the Milky Way's Galactic plane. This includes both the Faraday depth dispersion along a single line of sight, through a random walk in Faraday depth space, and differences between lines of sight. Strictly speaking, the Faraday depth distribution in NGC~891 applies to diffuse emission mixed with the plasma. Extragalactic sources behind the Milky Way probe lines of sight between Earth and the edge of the Milky Way along diverging lines of sight. With these caveats, a scale of $0.66\ \rm kpc$ corresponds to $\sim 4\degr$ at a distance of 10 kpc. The standard deviation of RM of extragalactic sources in the Galactic plane on the scale of a few degrees is in the order of $\sigma_\phi = 200\ \rm rad\ m^{-2}$ \citep[e.g.][]{vaneck2011,unger2024,shanahan2019}. This value somewhat underestimates the Faraday depth dispersion of an edge-on galaxy like NGC~891 because of our location inside the Milky Way and because lines of sight through the Milky Way are not parallel. For $\lambda^2 = 5 \times 10^{-3}\ \rm m^2$ (Figure~\ref{fig:fracpol2_lambda2}), we therefore expect for the complete disk $\sigma_\phi \lambda^2 \gtrsim 1$ in S-band, not even including the higher density of the WIM in NGC~891 \citep{rand1990}. 

The data suggest that $\sigma_\phi \lambda^2 \lesssim 1$, indicating that the diffuse polarized emission does not probe the entire disk, in analogy to the polarization horizon in the Milky Way \citep{uyaniker2003}. Also, Figure \ref{fig:AVG_RMprofileBvector}b shows a sign of the large-scale magnetic field with an RM amplitude of $\pm 22\ \rm rad\ m^{-2}$, which is much smaller than the expected Faraday depth of an edge-on disk. For the Milky Way we estimate a total Faraday depth $\Phi_0\ \approx 500\pm200\ \rm rad\ m^{-2}$ and $\sigma_\Phi \approx 200\ \rm rad\ m^{-2}$. If we fix the Faraday depth of the layer that we detect in polarization as $\Phi\approx70\ \rm rad\ m^{-2}$, then the depth of this layer, $\ell$, as a fraction of the radius of the disk, $R_d \approx 9.2\ \rm kpc$ (Figure \ref{fig:H_alpha_maps}a), can be estimated assuming a uniform disk with an axial magnetic field.
For lines of sight at normalized galactocentric radii projected in the plane of the sky, $b/R_d = 0.2$, $0.4$, $0.6$, and $0.8$, we obtain 
$\ell/R_d = 0.48$, $0.36$, $0.27$, and $0.18$, respectively. 

An independent estimate can be obtained from the Faraday depth dispersion. Assuming $\sigma_\Phi \approx 200\ \rm rad\ m^{-2}$ for the full disk, based on the Milky Way, and $\sigma_\phi \approx 100\ \rm rad\ m^{-2}$ for the polarized layer detected in our observations (Figure \ref{fig:fracpol2_lambda2}), we obtain $\ell/R_d = 0.50$, $0.46$, $0.40$, and $0.30$ for $b/R_d = 0.2$, $0.4$, $0.6$, and $0.8$, respectively. In this case the result depends on the inverse square of the adopted $\sigma_\phi$ for the complete disk.

Both estimates for the depth of the layer from which we receive detectable polarization are necessarily uncertain. Both indicate that we are observing a region with depth that is a considerable fraction of the radius of the radio disk. As such, we can expect that most H\,{\footnotesize II} regions visible in Figure~\ref{fig:H_alpha_maps} are within the region from which we detect polarized emission.

The lower fractional polarization in the midplane at short wavelengths is the result of unpolarized synchrotron emission from anywhere along the line of sight that contributes only to total intensity. \citet{beck2016magnetic} argued that the low fractional polarization in spiral arms of face-on galaxies is caused by a stronger random field in the arms. This should also be the case when a galaxy is observed edge-on.   

\subsection{Interpretation of Faraday depth components}

\begin{figure*}
    \centering

    \begin{minipage}[t]{0.48\textwidth}
        \centering
         \begin{overpic}[trim=10 10 10 10, clip, width=\linewidth]{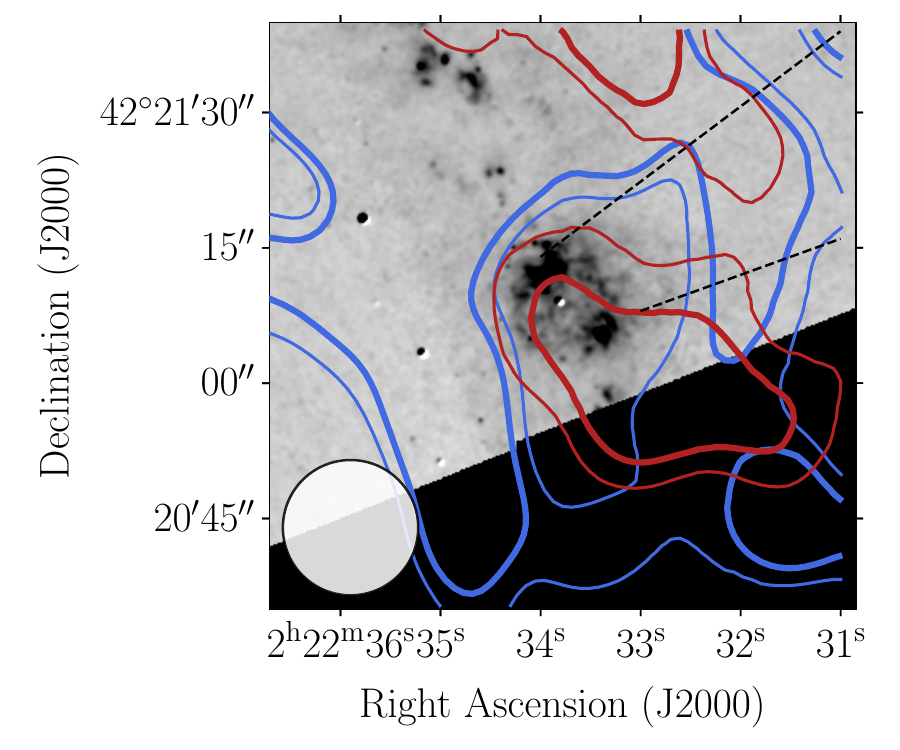}
            \put(30,78){\color{black}\setlength\fboxsep{1pt}\fcolorbox{black}{white}{\textbf{(a)}}}
        \end{overpic}
    \end{minipage}
    \begin{minipage}[t]{0.48\textwidth}
        \centering
        \raisebox{-0.01mm}{
        \begin{overpic}[trim=130 180 100 60, clip, width=0.6\linewidth]{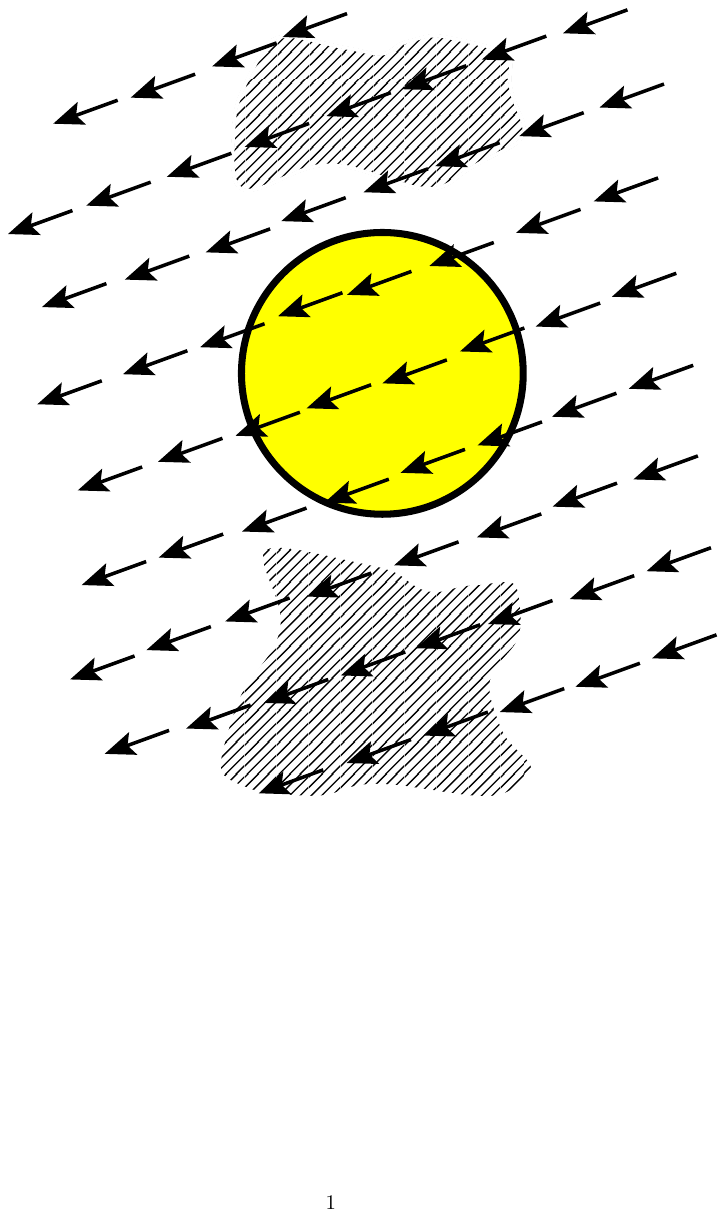}
            \put(0,89){\color{black}\setlength\fboxsep{1pt}\fcolorbox{black}{white}{\textbf{(b)}}}
        \end{overpic}}
    \end{minipage}

    \vspace{-12mm} 

    \begin{minipage}[t]{0.48\textwidth}
        \centering
        \begin{overpic}[trim=10 10 10 10, clip, width=\linewidth]{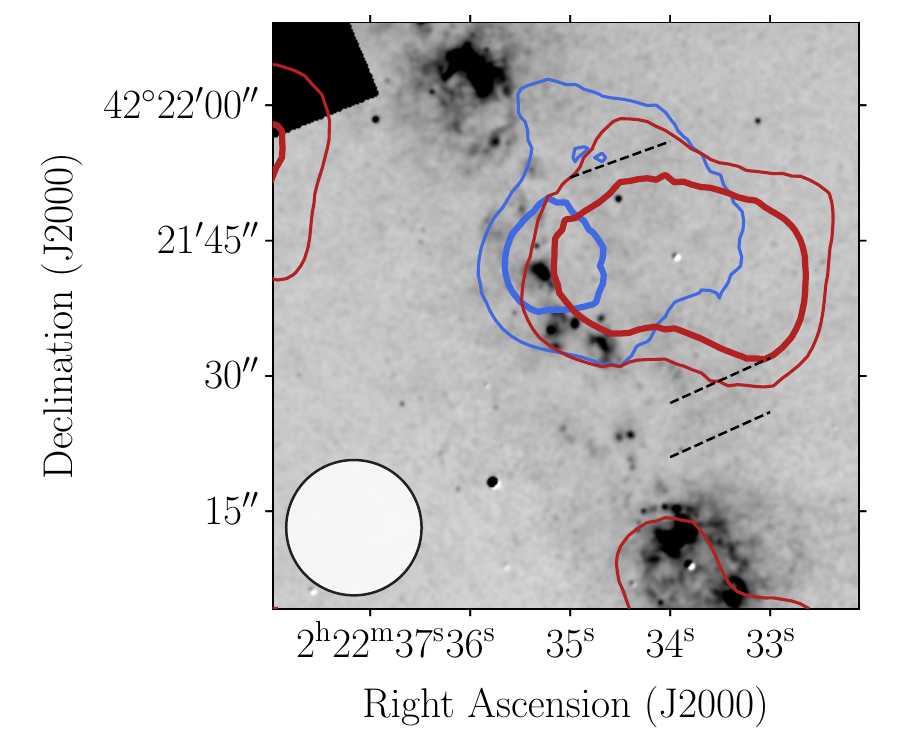}
            \put(30,78){\color{black}\setlength\fboxsep{1pt}\fcolorbox{black}{white}{\textbf{(c)}}}
        \end{overpic}
    \end{minipage}
    \begin{minipage}[t]{0.48\textwidth}
        \centering
        \raisebox{-20mm}{
        \begin{overpic}[trim=130 180 100 30, clip, width=0.8\linewidth]{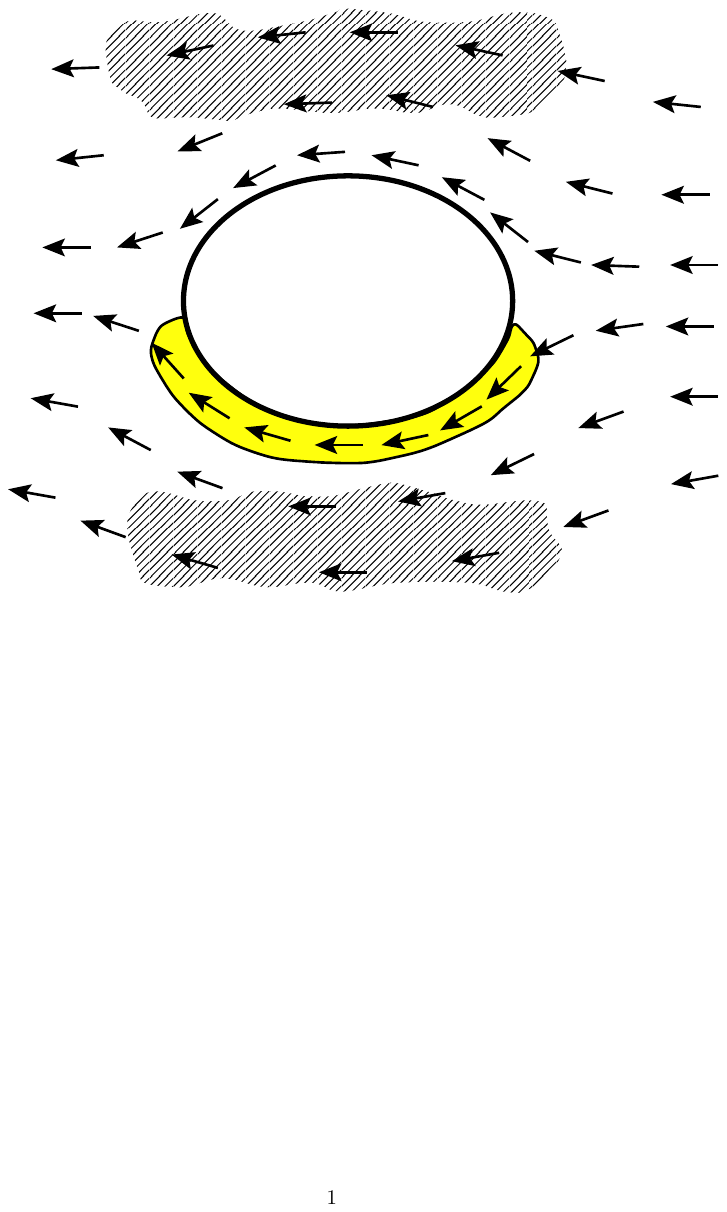}
            \put(10,82){\color{black}\setlength\fboxsep{1pt}\fcolorbox{black}{white}{\textbf{(d)}}}
        \end{overpic}}
    \end{minipage}
    \vspace{-18mm}

    \caption{Interpretation of Faraday depth spectra with multiple peaks. The case for H\,{\footnotesize II} regions 10 and 11 is shown in panels (a) and (b). Panel (a) shows in grayscale H$\alpha$ intensity from an archival image of the Hubble Space Telescope (HST). The dashed lines mark the approximate locations of filaments identified by \protect\citet{rossa2004hubble}. The black regions are outside the field of view. The blue contours show emission at the 6$\sigma$ and $8\sigma$ levels from Figure~\ref{fig:ngc891_fdf_spectrum_2peaks_UVIT}b, and the red contours show the same from Figure~\ref{fig:ngc891_fdf_spectrum_2peaks_UVIT}c. The shaded circle indicates the $15\arcsec$ (660 pc) beam. Panel (b) shows a top view of a non-emitting Faraday screen (yellow) with the observer in the galactic plane, looking from the bottom of the page. Black vectors indicate the magnetic field in the plane of the disk. Relevant regions emitting polarized synchrotron emission along the line of sight are schematically indicated by hatched areas. Panel (c) shows H\,{\footnotesize II} regions 12-14 with blue contours at $6\sigma$ and $8\sigma$ from Figure~\ref{fig:ngc891_fdf_spectrum_3peaks_UVIT}b and red contours from Figure~\ref{fig:ngc891_fdf_spectrum_3peaks_UVIT}c. Dashed lines flank H$\alpha$ filaments identified by \protect\citet{rossa2004hubble}. Panel (d) shows a top view with the observer at the bottom of the page, a non-emitting Faraday screen in yellow and relevant emitting regions as hatched areas.} 
    \label{fig:FD_comp}
\end{figure*}

Most locations show a single peak in the Faraday depth spectrum, with the spread of the RM Clean components indicating a Faraday depth range $\lesssim 100\ \rm rad\ m^{-2}$.

In a few locations we resolve multiple Faraday depth components in the Faraday depth spectrum (Figure~\ref{fig:ngc891_fdf_spectrum_3peaks_UVIT} and Figure~\ref{fig:ngc891_fdf_spectrum_2peaks_UVIT}). There is no unique configuration that gives rise to a Faraday depth spectrum with multiple peaks.  Figure~\ref{fig:FD_comp} offers interpretations based on the association with a super bubble and the assumption of a non-emitting Faraday screen with foreground and background diffuse polarized emission. The foreground and background emission regions are in reality much larger than shown. 

H\,{\footnotesize II} regions 10 and 11 are located in a larger shell seen in the Hubble Space Telescope H$\alpha$ image in Figure~\ref{fig:FD_comp}a, which is approximately the size of our beam. The red contours are associated with polarized emission in the background being displaced by $\sim 160\ \rm rad\ m^{-2}$ in Faraday depth by the plasma associated with the shell. A similar Faraday depth structure has been observed on smaller scales in the Milky Way \citep{mohammed2024}. It is possible that the bright H{\footnotesize II} regions 10 and 11 depolarize the background in a few regions smaller than the beam, while the Faraday rotation occurs in the more diffuse region. Figure~\ref{fig:H_alpha_maps} places the H\,{\footnotesize II} regions $\sim 4\ \rm kpc$ from the centre, but their far-UV emission indicates they are likely in front of most of the dust lane of NGC~891.  

The triple peak in Figure~\ref{fig:ngc891_fdf_spectrum_3peaks_UVIT} requires a more complex configuration. We have dismissed the possibility of a background radio source, because the structure is well resolved and there is no evidence for a Stokes $I$ counterpart in the Stokes $I$ contours. The two extreme Faraday depth components are slightly offset with respect to each other, between H$\alpha$ filaments that appear to indicate an outflow from H\,{\footnotesize II} regions 12-14 \citep{rossa2004hubble}. 

Figure~\ref{fig:FD_comp}d illustrates how an asymmetric shell with density enhancement on one side, for example resulting from a density gradient, can act as a Faraday screen that creates two Faraday depth components with a spatial displacement comparable to the diameter of the shell. At our angular resolution, the Faraday depth components appear to overlap on the sky, but at higher resolution they should be observed separately. This model requires an asymmetric superbubble and the observer's line of sight at a large angle with the large-scale magnetic field. The latter is supported by the location of the H\,{\footnotesize HII} regions in Figure~\ref{fig:H_alpha_maps}a. 

\citet{maconi2025} presented 3-dimensional magneto-hydrodynamic simulations of a supernova driven interstellar medium with superbubbles with diameter $\sim 200\ \rm pc$, that display large-scale Faraday depth structure with amplitude $\sim 200\ \rm rad\ m^{-2}$. The larger Faraday depth range observed here can be mostly attributed to a longer path length through a bubble that is $\sim 4$ times larger. The brightest Faraday depth component in Figure~\ref{fig:ngc891_fdf_spectrum_3peaks_UVIT}a is identified with polarized emission in the foreground.

In this geometry, the superbubble may be a non-emitting Faraday screen with foreground and background diffuse polarized emission, provided that the shell does not completely depolarize the emission in the background. The extreme Faraday depth components would arise from the two sides of the shell that would be observed separately at higher angular resolution. The bright Faraday depth component near zero Faraday depth is foreground emission, observed across most of the galaxy.

In summary, our analysis suggests that the enhanced star formation and its associated diffuse X-ray emission occurs in a localized region in the outer earth-facing disk of NGC~891. The enhanced polarized emission is associated with this structure, possibly as part of a large superbubble. The modest wavelength dependence of the depolarization of synchrotron emission in the plane of the disk suggests that the low-latitude polarization in an edge-on spiral galaxy in general may originate in the outer disk, with most of the emission being depolarized by line-of-sight depolarization within the galaxy.  

\section{Conclusion} \label{sec:Conclusion}
We present new VLA S-band observations combined with previously published C-band polarization observations of NGC~891, examining the structure of its radio continuum halo using Rotation Measure Synthesis. Our analysis leads to the following results and conclusions: 
\begin{itemize}
    \item A region on the north-east side of the galaxy shows high polarized intensity, accompanied by diffuse X-ray emission and a local excess in Stokes $I$. This region was previously found to contain multiple superbubbles, and it appears on the Earth-facing side of NGC~891, near the edge of the starforming/radio continuum disk. The morphology and multiwavelength properties suggest a large superbubble powered by a relatively young star cluster. 
    \item The RM map of NGC~891, corrected for foreground Faraday rotation shows a smooth transition along the major axis, between the north-east (mean RM = $4.50\ \rm rad\ m^{-2}$) and the south-west (mean RM = $48.42\ \rm rad\ m^{-2}$). A compact region of high positive RM ($\sim250\ \rm rad\ m^{-2}$) toward the galaxy’s center is identified.
    \item Across both disk and halo, the fractional polarization decreases toward the midplane but varies only weakly with wavelength. This indicates that the region from which polarized radio emission is detected has small internal Faraday dispersion and small differential Faraday rotation within the beam from any plasma in the foreground. We surmise that most of the detected polarized emission originates on the Earth-facing side of NGC~891. The low fractional polarization in the midplane is therefore primarily due to unpolarized synchrotron emission.
    \item Most of the bright H\,{\footnotesize II} regions in the northeast side of NGC~891 are closer to the center than the H\,{\footnotesize II} regions in the southwest. The H\,{\footnotesize II} regions on both sides of the galaxy probably do not follow a common logarithmic spiral arm. In addition, we identified a faint, isolated H\,{\footnotesize II} region at a galactocentric radius of $16.90\ \rm kpc$ with $\rm H\alpha$ and far-UV counterparts, indicating recent star formation well outside the main star forming disk. 
\end{itemize}

In the future, we will extend the analysis to include L-band data. This will improve the resolution in Faraday depth but this will be more useful only in selected areas where significant polarized intensity exists at these longer wavelengths. 

\begin{acknowledgments}
The authors thank the anonymous referee for comments that helped improve the manuscript. The National Radio Astronomy Observatory and Green Bank Observatory are facilities of the U.S. National Science Foundation operated under cooperative agreement by Associated Universities, Inc. This paper uses observations made with the NASA/ESA Hubble Space Telescope, and obtained from the Hubble Legacy Archive, which is a collaboration between the Space Telescope Science Institute (STScI/NASA), the Space Telescope European Coordinating Facility (ST-ECF/ESAC/ESA) and the Canadian Astronomy Data Centre (CADC/NRC/CSA). NP acknowledges the use of AI-assisted tools for language refinement and code verification
during manuscript preparation. 
JMS acknowledges the support of the Natural Sciences and Engineering Research Council of Canada (NSERC), 2019-04848, and a generous private donor of a substantial server for data processing and simulations. 
J.T.L. acknowledges the financial support from the science research grants from the National Science Foundation of China (NSFC) through grants 12321003 and 12273111, and the China Manned Space Program with grant no. CMS-CSST-2025-A10 and CMS-CSST-2025-A04.
S. Ranasinghe acknowledges financial support from the Canadian Space Agency for his valuable assistance with the UVIT data analysis.
PK is partially supported  by the BMBF project 05A23PC1 for D-MeerKAT III.
TW acknowledges financial support from the grant CEX2021-001131-S funded by MICIU/AEI/ 10.13039/501100011033, from the coordination of the participation in SKA-SPAIN, funded by the Ministry of Science, Innovation and Universities (MICIU).
This research has made use of data obtained from the Chandra Data Archive provided by the Chandra X-ray Center (CXC). This publication uses the data from the AstroSat mission of the Indian Space Research Organisation (ISRO), archived at the Indian Space Science Data Centre (ISSDC). This research has made use of the NASA/IPAC Extragalactic Database (NED), which is operated by the Jet Propulsion Laboratory, California Institute of Technology,
under contract with the National Aeronautics and Space Administration. We acknowledge the usage of the HyperLeda database.
This study used the following packages: CosmosCanvas \citep{english2024cosmoscanvas}.
\end{acknowledgments}

\bibliography{references}{}
\bibliographystyle{aasjournalv7}

\end{document}